# The Sky Distribution and Magnitudes of Starlink Satellites by the Year 2027

Anthony Mallama

anthony.mallama@gmail.com

2022 September 1

Abstract

Visual magnitudes and sky coordinates are projected for the full constellation of Starlink satellites. The results are presented in the form of sky maps and numerical tables. Observer latitudes from the equator to 60$^o$ are considered. The solar elevations include -12$^o$ (the end of nautical twilight), -18$^o$ (the end of astronomical twilight) and -30$^o$.

1. Introduction

The largest group of spacecraft currently in orbit is the constellation of Starlink satellites. They presently number over 2,000 and are projected to total nearly 12,000 by year 2027. Professional astronomers are already experiencing interference from Starlink spacecraft when making celestial observations (Mroz et al., 2022) and amateur stargazers are concerned as well for both scientific and aesthetic reasons (Mallama and Young, 2021). The International Astronomical Union has established a Centre for the Protection of Dark and Quiet Skies from Satellite Constellation Interference to address this problem. A reading list of papers on satellite constellations and their impact is provided in Appendix A.

This paper forecasts the distribution and brightness of Starlink satellites in the sky after the entire constellation is in place. The method is to compute the azimuth, elevation and range for each satellite visible at ground locations from the equator to latitude 60°. These coordinates are based on spacecraft orbital elements and they allow for computation of the apparent magnitude. Brightness is calculated from a phase curve equation derived from visual magnitudes.

Section 2 lists the spacecraft orbital elements and describes the method of computing satellite topocentric coordinates. The method of determining whether a satellite is in sunlight or in the Earth's shadow is also explained. Section 3 describes the phase functions for three Starlink satellite designs. These curves characterize brightness versus the phase angle which is measured at the satellite between the Sun and the Earth. Then the equation for computing apparent magnitudes is presented. More details about the Starlink phase functions can be found in Appendix B. Section 4 gives examples of resulting sky maps which plot satellite locations by azimuth, elevation and magnitude. Sample tables of numerical results are also provided. Comprehensive sets of map and numerical results for a wide range of observer latitude and solar distances below the horizon are given in Appendix C. Section 5 considers the limitations and uncertainties in this study. Section 6 presents the conclusions and compares the satellite number density to an earlier forecast made by McDowell (2020).

2. Satellite orbits and celestial coordinates

Starlink satellites will occupy in 8 'shells' at altitudes ranging from 336 to 570 km as listed in Table 1. There are several orbital planes for each shell and a number of satellites in each plane. For this study the ascending nodes of the planes for each shell are distributed evenly over 2 π radians. Likewise, the orbital anomaly for each satellite in a plane is 2 π divided by the number of satellites in that plane. The geocentric coordinates of satellites are determined from the elements including the anomalies. Examples of these coordinates for the satellites in medium inclination and high inclination shells are shown in Figures 1 and 2, respectively.

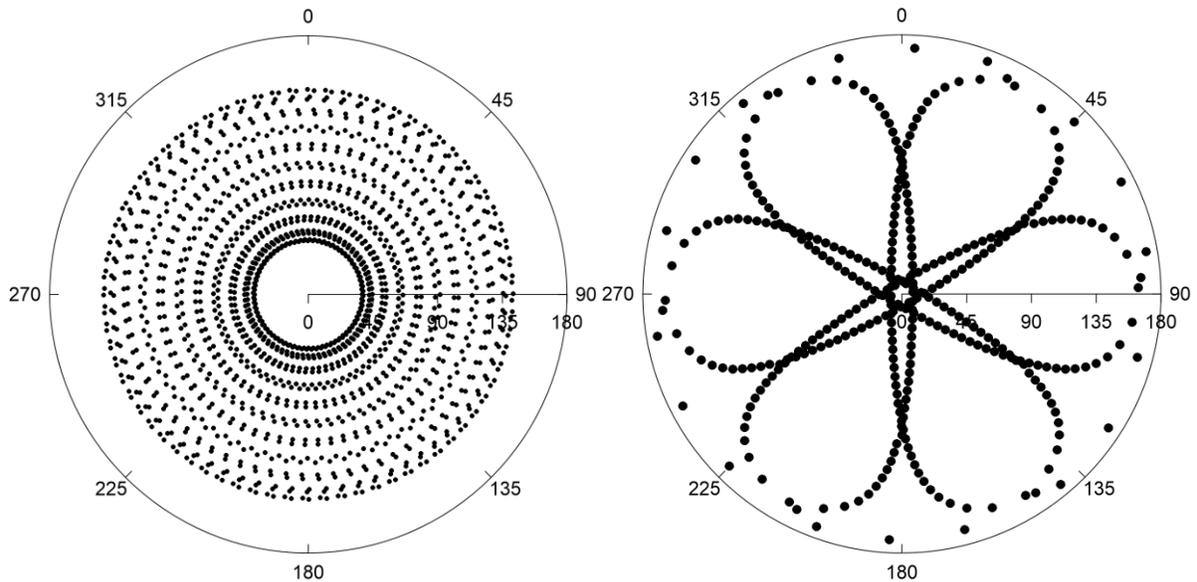

Figure 1 (left). The distribution of satellites in shell 1 with orbital inclination 53.0 degrees is shown as a polar plot with the radial axis extending from 0 degrees in the direction of the north pole of the Earth's axis to 180 in the opposite direction. Satellite positions are concentrated toward the equator and away from the poles.

Figure 2 (right). The distribution of satellites in shell 4 with orbital inclination 97.6 degrees is illustrated. Satellite positions are more concentrated toward polar latitudes than in Figure 1.

Table 1. Orbital information

```
Shell            1      2      3      4      5      6      7      8
Altitudes        550.   540.   570.   560.   560.   336.   341.   346.
Inclinations     53.0   53.2   70.0   97.6   97.6   42.0   48.0   53.0
Planes           72     72     36     6      4      50     50     50
Sats_per_plane   22     22     20     58     43     50     50     50

The information in Table 1 is taken from
https//en.wikipedia.org/wiki/Starlink as retrieved on 2022 July 15. 'Planes'
here is 'count' in the source document, and 'sats_per_plane' here is
'sat_per' in the source. The number of planes and satellites per plane for
shells 6 through 8 was not listed in the source document, so each is taken to
be approximately the square root of 'satellites' there.
```

The satellite azimuth and elevation at a selected ground location are determined by transforming its geocentric coordinates to topocentric values at the latitude and longitude on the ground. Examples of the satellite azimuths and elevations shown in Figures 3 and 4 illustrate how the distribution changes with observer latitude.

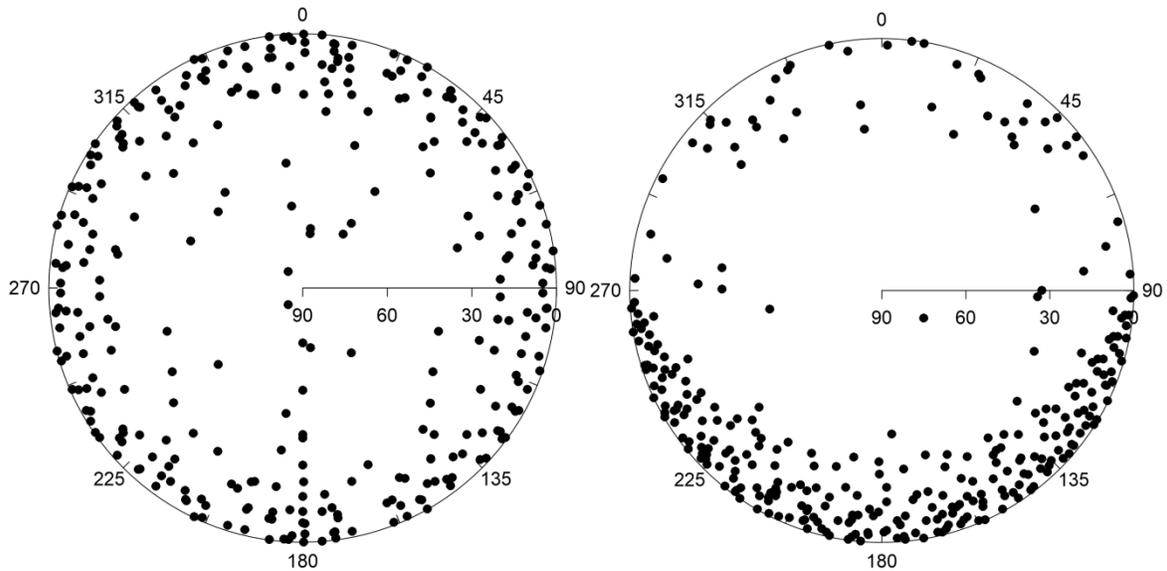

Figure 3 (left). The distribution of satellites above the horizon as seen from the equator is plotted. Density increases with distance from zenith but the azimuthal distribution is practically uniform.

Figure 4 (right). The satellite distribution as seen from latitude 60° north is heavily concentrated toward the south (azimuth 180) because most orbital inclinations are less than that latitude.

Satellites may be hidden from sunlight by the Earth's shadow in which case they do not interfere with optical astronomy. The shadow limit at a shell's orbital altitude corresponds to an angular radius measured at center of the Earth relative to a line extending from the Sun through that geocenter location. The radius values are given by the arc-cosine of the Earth's radius divided by the distance of a satellite from the geocenter. These shadow radii range from about 67° for the highest altitude satellites to 72° for the lowest. Figures 7 and 8 in Section 4 illustrate how the Earth's shadow eclipses satellites over a larger area of sky as the Sun's distance below the horizon grows.

3. Phase function and satellite brightness

Phase curves describe spacecraft brightness as a function of the angle measured at the satellite between the Sun and the observer. The chassis of a Starlink spacecraft is shaped like a flat panel and its orientation is nominally perpendicular to the nadir-zenith direction. Additionally, there is a large solar array attached at a right angle to the zenith-facing side of the panel and it projects upward. These components are shown schematically in Figure 5 along with their orientation in space.

Sunlight scatters from Starlink spacecraft in a manner that differs sharply from that of round bodies like the Moon and planets. Those celestial bodies preferentially reflect light back in the direction from which it originated. Starlink spacecraft, on the other hand, reflect sunlight both backwards and forwards. The forward scattering is likely due to a near-specular reflection from the flat nadir side.

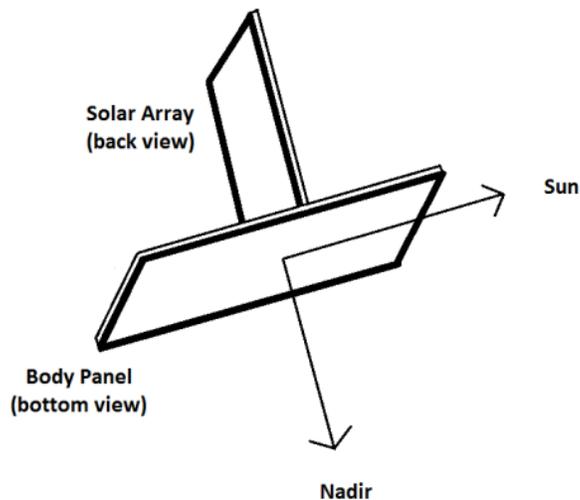

Figure 5. This schematic of the Starlink spacecraft illustrates its shape and orientation relative to the Sun and Earth.

There have been two changes to the detailed design of Starlink satellites, so far. The Original Design was the brightest and about 400 spacecraft (https://planet4589.org/space/stats/star/starstats.html, derived from data retrieved 2022 June 15) were launched before it was modified in response to the concerns of astronomers. The manufacturer, SpaceX, added a sunshade to help reduce luminosity and about 1,100 of this new model, called VisorSat, were launched (ibid). VisorSats were typically 1.3 magnitudes fainter than the Original Design (Mallama, 2021a) at operational altitudes. However, when SpaceX upgraded the communication system for Starlink from radio to laser in 2021 the sunshades had to be dropped from the design (Witze, 2022). In order to compensate for removing the shades SpaceX began development of reflective materials for the nadir side of the flat panel to direct sunlight away from observers beneath the spacecraft. The new design occurred at about serial number 3000 and so this latest model is referred to as Series-3000 herein. This latest design of Starlink is typically about 0.5 magnitudes brighter than VisorSats although it is still approximately 0.8 magnitudes fainter than the Original Design (Mallama, 2022).

Besides their different characteristic luminosities each Starlink design has its own distinct phase function. Figure 6 shows that the response for the Original Design is relatively flat, while that of VisorSat is strongly curved and that for Series-3000 is intermediate between the two. Since the phase function for Series-3000 lies between the two others and because this model represents the current Starlink design, the magnitudes in this paper rely on its phase function as indicated by Eq. 1,

$$v = 3.817 + 0.06970 \, \phi - 0.0003999 \, \phi^2 + 5 \log (d / 1000)$$

Eq. 1

where *v* is the apparent visual magnitude, *φ* is the phase angle in degrees, and *d* is the distance from the satellite to the observer in km. The distance term accounts for the inverse square law of light.

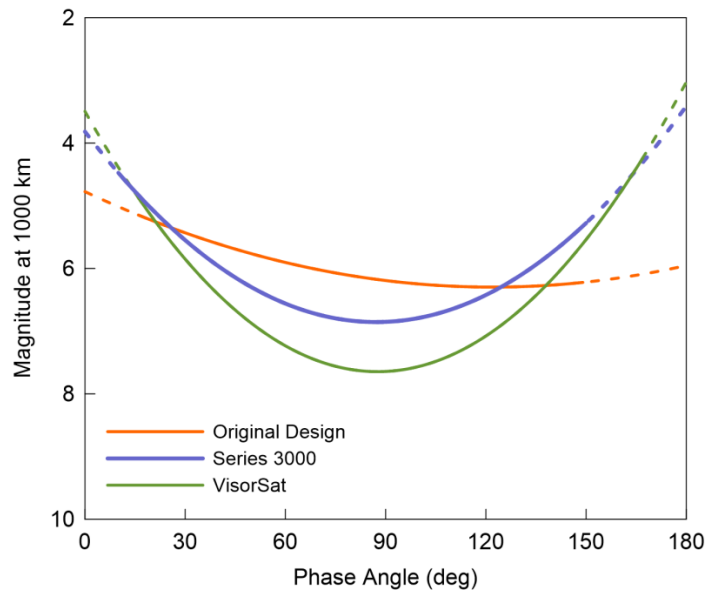

Figure 6. Phase functions corresponding to the three Starlink designs for spacecraft at their operational altitudes. The visual magnitudes used to derive these functions are close to the V-band of the Johnson-Cousins photometric system of electronic measurements. Dashed lines are extrapolations beyond the observed range; satellites are rarely seen at these extreme phase angles.

The phase function accounts for some of the apparent brightness of Starlink satellites. However, their complex shapes make the luminosity sensitive to spacecraft attitude (yaw, pitch and roll) in a way that is not yet generally understood. Brightness variations due to attitude appear as scatter in an empirically derived phase function diagram. A study of Starlink magnitudes (Mallama, 2021b) indicated that this scatter was about 0.4 magnitudes after observational uncertainties were removed. Therefore, the apparent magnitudes computed in this paper have that amount of scatter applied. The Box-Muller algorithm for normal statistical distributions as implemented here:

[https://masuday.github.io/fortran_tutorial/random.html](https://masuday.github.io/fortran_tutorial/random.html) is employed. More details about the Starlink phase function and the scatter are provided in Appendix B.

4. Distribution in the sky and apparent magnitudes

The geometric and photometric aspects of Starlink satellites discussed in the previous two sections allow for simulation of their appearance in the sky. Examples of sky maps showing distributions and magnitudes as well as numerical output are presented in this section. The statistical results include satellite counts by magnitude, by elevation and by quadrant of the sky, as well as the number of satellites in sunlight and in shadow.

Maps for an observer at the terrestrial equator and with the Sun on the celestial equator are shown as Figures 7 and 8, while corresponding numerical data is given in Tables 2 and 3. At the end of astronomical twilight when the Sun is 18° below the horizon (Figure 7 and Table 2) satellites are visible in all quadrants of the sky; 89 of them are brighter than magnitude 7, and 21 of those are above 30° elevation. When the Sun has descended to 30° below the horizon (Figure 8 and Table 3) there no satellites visible in the quadrants opposite the solar azimuth; only 19 satellites are brighter than magnitude 7 and none of those are above 30° elevation.

Similar maps and tables for observer latitudes from the equator to 60° are presented in Appendix C. Results are given for the dates of solstice and equinox, and for solar distances of 12°, 18° and 30° below the horizon. The distribution and magnitudes of satellites is discussed in the Section 6.

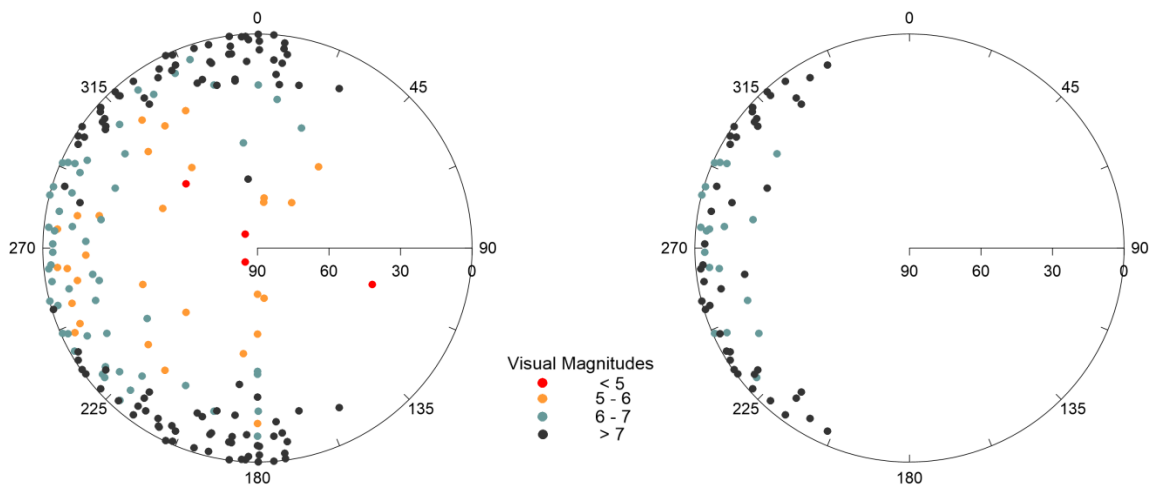

Figure 7 (left). The distribution and brightness of Starlink satellites for an observer at the equator on a date of equinox with the Sun 18° below the horizon at azimuth 270° is plotted.

Figure 8 (right). The Sun is 30° below azimuth 270°. This plot along with Figure 7 illustrates how the Earth's shadow progressively hides more satellites as the Sun descends further below the horizon. They also demonstrate how the forward-scattering phase curve enhances the brightness of satellites that are aligned with the solar azimuth.

Table 2. Satellite locations and magnitudes corresponding to Figure 7.

```
Latitude    0.0,   solar az. 270.0,   solar el. -18.0,   solar dec.   0.0
           ----------------------------- Quadrant ----------------------------
Magni-     Northwest       Southwest       Southeast       Northeast       All Quadrants
tude       30+    30-      30+    30-      30+    30-      30+    30-      30+    30-    Any
 < 5         2      0        1      0        1      0        0      0        4      0      4
 5 - 6       2      7        3      9        3      1        4      0       12     17     29
 6 - 7       1     23        1     24        2      2        1      2        5     51     56
 > 7         1     41        1     36        0     15        0     13        2    105    107
Sunlit       6     71        6     69        6     18        5     15       23    173    196
Shadow       0      0        0      0        2     56        2     59        4    115    119
All          6     71        6     69        8     74        7     74       27    288    315
```

Table 3. Satellite locations and magnitudes corresponding to Figure 8.

```
Latitude    0.0,   solar az. 270.0,   solar el. -30.0,   solar dec.   0.0
           ----------------------------- Quadrant ----------------------------
Magni-     Northwest       Southwest       Southeast       Northeast       All Quadrants
tude       30+    30-      30+    30-      30+    30-      30+    30-      30+    30-    Any
 < 5         0      0        0      0        0      0        0      0        0      0      0
 5 - 6       0      0        0      0        0      0        0      0        0      0      0
 6 - 7       0     11        0      8        0      0        0      0        0     19     19
 > 7         0     21        0     23        0      0        0      0        0     44     44
Sunlit       0     32        0     31        0      0        0      0        0     63     63
Shadow       6     39        6     38        8     74        7     74       27    225    252
All          6     71        6     69        8     74        7     74       27    288    315
```

5. Limitations of this study

Several factors place limits on the results presented here. Most importantly, the future of the Starlink constellation is uncertain. The source of the orbital data listed in Table 1 does not give complete Keplerian elements and some assumptions had to be made as stated there. Furthermore, since only a fraction of the satellites have actually been launched the actual configuration and number of the Starlink satellites in space may turn out to be different. Also, the projected completion date for launching all of the nearly 12,000 satellites may not be met.

Another issue is the phase curve. The magnitudes in this study are computed from the curve for Series-3000 satellites. However, there have already been three Starlink designs and each has had its own distinctive reflection characteristics and phase curve. Future design modifications would likely result in new phase functions.

Besides changes to the shape of Starlink satellites, SpaceX CEO Elon Musk has stated in this video https://youtu.be/XP5k3ZzPf_0?t=561 that the future Starlink 2.0 will be a larger spacecraft with a length of 7 meters. This increased size may reflect more light and appear brighter in the sky.

Finally, SpaceX has requested approval to launch 30,000 satellites in addition to the 12,000 already approved. So, the sky could become much more crowded with bright satellites than this study projects.

6. Conclusions

The sky maps and numerical results in Section 4 and in Appendix C indicate the impact of Starlink satellites on celestial observation by the year 2027. During astronomical twilight hundreds of spacecraft brighter than magnitude 7 will be visible from any one location. However, most of those will be within 30$^o$ of the horizon and only tens of satellites will be seen at higher elevations. The number of visible satellite diminishes rapidly after astronomical twilight ends.

Additionally, spacecraft that are aligned with the solar azimuth and those opposite that azimuth tend to be brighter because of forward and backward scattering of light. Finally, the density of satellites is greater toward celestial south for an observer north of the terrestrial equator and vice-versa due to the geocentric distribution of satellites.

Finally, the results obtained here are compared with two previous studies. First, McDowell (2020) forecast "a latitude-dependent areal number density of between 0.005 and 0.01 objects per square degree at airmass <2". The corresponding numbers from this study range from 0.003 at the equator to 0.006 at 50$^o$ latitude. These smaller densities are due lower orbital altitudes for satellites in shells 2 through 5. At the time of McDowell's study these heights were expected to range from 1110 to 1325 km but now they are from 540 to 560. Second, Lawler et al. (2022) predict that latitudes near 50$^o$ will experience the worst satellite interference. This analysis finds that the number density increases to that latitude as noted above, and then it falls off sharply by latitude 60$^o$. So, these findings agree.

Appendix A. Reading list of papers on satellite constellation interference

These papers are grouped by subject matter as follows: impacts on observations, working group reports and conference material, satellite brightness, mitigation strategies and general. Titles are given here to indicate content while full citations are provided in the Reference section.

Appendix A-1. Impacts on celestial observations

Gallozzi et al. 2020. Concerns about ground based astronomical observations: a step to safeguard the astronomical sky.

Hainaut and Williams, 2020. Impact of satellite constellations on astronomical observations with ESO telescopes in the visible and infrared domains.

McDowell, 2020. The low Earth orbit satellite population and impacts of the SpaceX Starlink constellation.

Mroz et al. 2022. Impact of the SpaceX Starlink satellites on the Zwicky Transient Facility survey observations.

Williams et al. 2021b. A report to ESO Council on the impact of satellite constellations.

Walker et al. 2020a. Impact of satellite constellations on optical astronomy and recommendations toward mitigations.

Williams et al. 2021a. Analysing the impact of satellite constellations and ESO's role in supporting the astronomy community.

Appendix A-2. Working group reports and conference proceedings

Hall et al. 2021. SatCon2 Working Group Reports.

Walker and Benvenuti, 2022. Dark and Quite Skies for Science and Society II.

Walker et al. 2020b. Dark and quiet skies for science and society.

Otarola et al. 2020. Draft Report of the Satellite Observations Working Group.

Appendix A-3. Satellite brightness

Halferty et al. 2022. Photometric characterization and trajectory accuracy of Starlink satellites: implications for ground-based astronomical surveys.

Hossein et al. 2022. Photometric characterization of Starlink satellite tracklets using RGB filters.

Krantz et al. 2021. Characterizing the all-sky brightness of satellite mega-constellations and the impact on astronomy research.

Mallama, 2021a. The brightness of VisorSat-design Starlink satellites.

Mallama, 2021b. Starlink satellite brightness – characterized from 100,000 visible light magnitudes.

Mallama, 2022. Newest Starlink satellites have gotten brighter again.

Tregloan-Reed et al. 2020. First observations and magnitude measurement of Starlink's Darksat.

Appendix A-4. Mitigation strategies

Tyson et al. 2020. Mitigation of LEO satellite brightness and trail effects on the Rubin Observatory LSST.

Appendix A-5. General audience

Mallama and Young, 2021. The satellite saga continues.

Witze, 2022. 'Unsustainable': how satellite swarms pose a rising threat to astronomy.

Appendix B. Phase functions

The material in this appendix supplements the discussion of phase functions in Section 3. The empirically derived phase functions for Original Design, VisorSat and Series-3000 spacecraft plotted in Figure X are shown along with the observations in Figures B-1, B-2 and B-3.

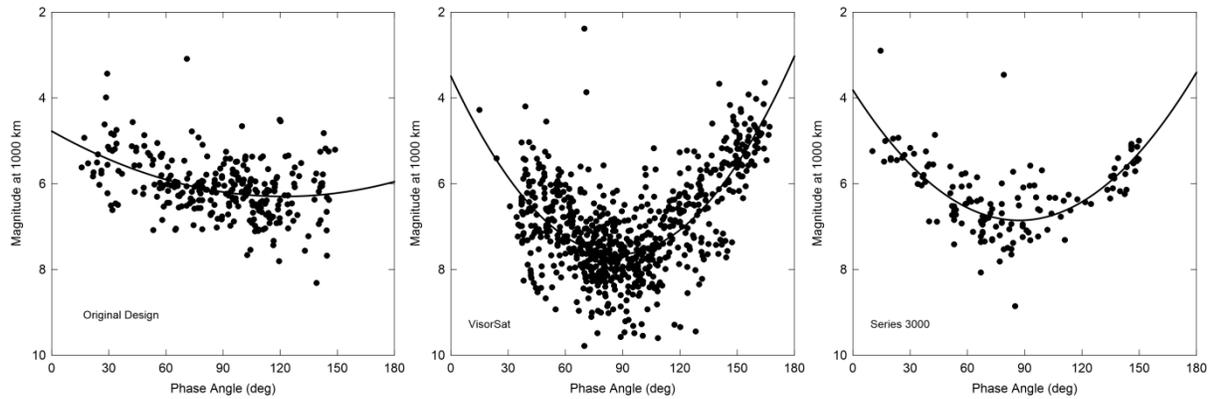

Figures B-1, B-2 and B-3 (left to right). The observed and modeled phase curves for Original Design, VisorSat and Series-3000 Starlink satellites are compared. The scatter is mostly due to variations in spacecraft attitude (yaw, pitch and roll).

A study of Starlink magnitudes measured electronically (Mallama 2021b) revealed that phase functions can change with time as illustrated in Figure B-4. This variation may have resulted from systematic changes to spacecraft attitude.

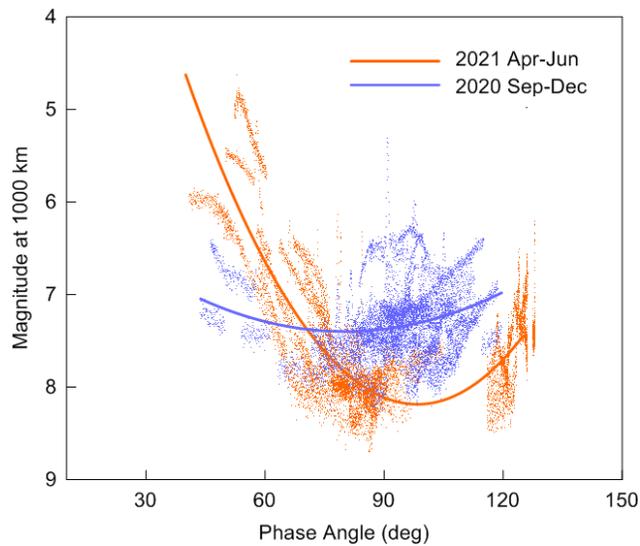

Figure B-4. Two markedly different VisorSat phase functions separated by several months are illustrated.

The plots in Figures B-5 and B-6 demonstrate moderate sensitivity of magnitudes in the sky to phase function by comparing Series-3000 to VisorSat. The magnitudes for Series-3000 are somewhat brighter at high elevations. This result is consistent with the phase curves in B-2 and B-3 where VisorSats are fainter around phase angle 90°. At low elevations the magnitudes are more nearly equal.

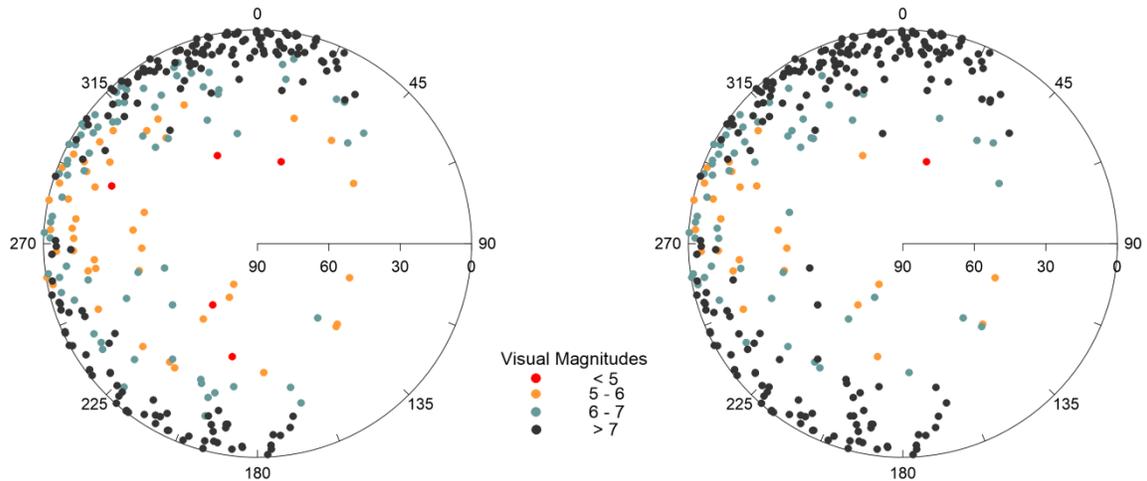

Figures B-5 (left) and B-6 are for an observer at 30° latitude with the Sun 18° below the horizon at azimuth 281°. B-5 is for the Series-3000 phase function and B-6 is for VisorSat.

The phase curves were derived from visual magnitudes observed by Jay Respler at latitude 40.330 and longitude 74.445 west, and by the author at 38.982 and 76.763. They are reported at http://www.satobs.org/seesat/index.html.

Appendix C. Sky maps and numerical tables for latitudes from 10° to 60°

Results for observer latitudes 0° through 60° are presented in appendix sections C-0 through C-6. Each section is divided into subsections α, β and γ for the Sun at declinations -23.4° (winter solstice), 0.0° (equinox) and +23.4° (summer solstice). The subsections contain sky maps and numerical results for the Sun's distance below the horizon of 12° (end of nautical twilight), 18° (end of astronomical twilight) and 30°, respectively. For example, in Appendix C-0-α the number '0' is for latitude 0°, the 'α' is for solar declination -23.4°, and the results include solar zenith angles of 102°, 108° and 120°. The maps and numerical results for southern observer latitudes are the same as their northern counterparts with two trivial exceptions; the sky quadrants on each side of the east-west line are reversed and the solar declinations are switched on the solstice dates.

Appendix C-0-α. Latitude 0°, solar declination -23.4°

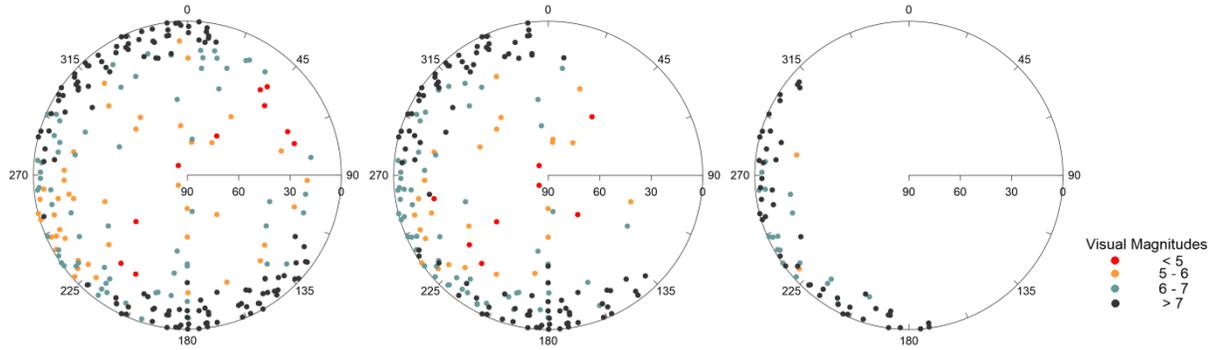

Figure C-0-α. The distribution and brightness of Starlink satellites is illustrated for an observer at the equator on a date when the Sun is at declination -23.4°. From left to right the Sun's distance below the western horizon is 12° (end of nautical twilight), 18° (end of astronomical twilight) and 30°.

Table C-0-α. Latitude 0°, solar declination -23.4°

```
Latitude    0.0,   solar az. 246.0,   solar el. -12.0,   solar dec. -23.4
           ---------------------------- Quadrant ----------------------------
Magni-     Northwest      Southwest      Southeast      Northeast      All Quadrants
tude       30+    30-     30+    30-    30+    30-     30+    30-    30+    30-    Any
 < 5        1      0       1      2      0      0       1      5      3      7      10
 5 - 6      3      5       2     20      3      6       4      1     12     32      44
 6 - 7      2     16       3     27      5      6       2     10     12     59      71
 > 7        0     50       0     20      0     43       0     10      0    123     123
Sunlit      6     71       6     69      8     55       7     26     27    221     248
Shadow      0      0       0      0      0     19       0     48      0     67      67
All         6     71       6     69      8     74       7     74     27    288     315

Latitude    0.0,   solar az. 245.3,   solar el. -18.0,   solar dec. -23.4
           ---------------------------- Quadrant ----------------------------
Magni-     Northwest      Southwest      Southeast      Northeast      All Quadrants
tude       30+    30-     30+    30-    30+    30-     30+    30-    30+    30-    Any
 < 5        1      0       2      3      1      0       1      0      5      3       8
 5 - 6      3      2       2     11      3      0       4      0     12     13      25
 6 - 7      2     14       2     33      3      4       0      1      7     52      59
 > 7        0     46       0     22      1     28       0      1      1     97      98
Sunlit      6     62       6     69      8     32       5      2     25    165     190
Shadow      0      9       0      0      0     42       2     72      2    123     125
All         6     71       6     69      8     74       7     74     27    288     315

Latitude    0.0,   solar az. 242.7,   solar el. -30.0,   solar dec. -23.4
           ---------------------------- Quadrant ----------------------------
Magni-     Northwest      Southwest      Southeast      Northeast      All Quadrants
tude       30+    30-     30+    30-    30+    30-     30+    30-    30+    30-    Any
 < 5        0      0       0      0      0      0       0      0      0      0       0
 5 - 6      0      1       0      1      0      0       0      0      0      2       2
 6 - 7      0      2       0     16      0      0       0      0      0     18      18
 > 7        0     14       0     29      0      3       0      0      0     46      46
Sunlit      0     17       0     46      0      3       0      0      0     66      66
Shadow      6     54       6     23      8     71       7     74     27    222     249
All         6     71       6     69      8     74       7     74     27    288     315
```

Appendix C-0-β. Latitude 0°, solar declination 0.0°

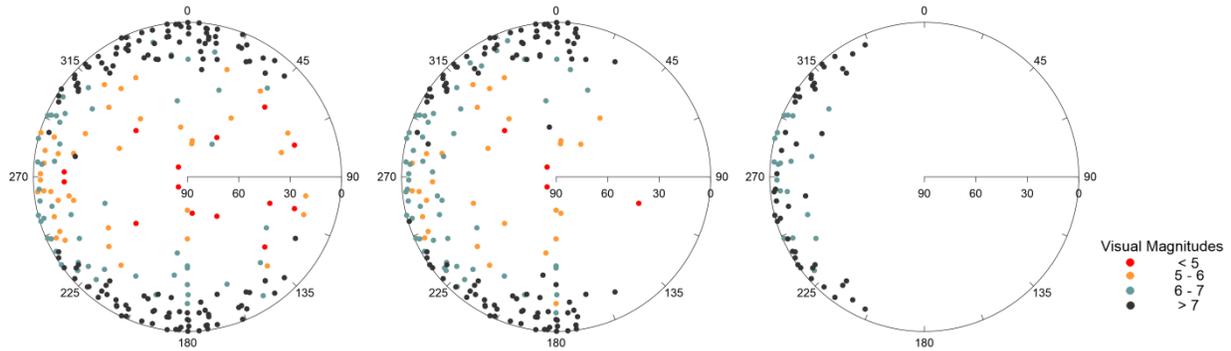

Figure C-0-β. The distribution and brightness of Starlink satellites is illustrated for an observer at the equator on a date when the Sun is at declination 0.0°. From left to right the Sun's distance below the western horizon is 12° (end of nautical twilight), 18° (end of astronomical twilight) and 30°.

Table C-0-β. Latitude 0°, solar declination 0.0°

```
Latitude   0.0,   solar az. 270.0,  solar el. -12.0,  solar dec.   0.0
           ---------------------------- Quadrant ----------------------------
Magni-     Northwest     Southwest     Southeast     Northeast      All Quadrants
tude       30+    30-    30+    30-    30+    30-    30+    30-    30+    30-    Any
 < 5        2      1      2      1      3      2      1      2      8      6     14
 5 - 6      3     12      1     13      2      3      4      3     10     31     41
 6 - 7      1     15      3     19      3      7      2      4      9     45     54
 > 7        0     43      0     36      0     27      0     28      0    134    134
Sunlit      6     71      6     69      8     39      7     37     27    216    243
Shadow      0      0      0      0      0     35      0     37      0     72     72
All         6     71      6     69      8     74      7     74     27    288    315

Latitude   0.0,   solar az. 270.0,  solar el. -18.0,  solar dec.   0.0
           ---------------------------- Quadrant ----------------------------
Magni-     Northwest     Southwest     Southeast     Northeast      All Quadrants
tude       30+    30-    30+    30-    30+    30-    30+    30-    30+    30-    Any
 < 5        2      0      1      0      1      0      0      0      4      0      4
 5 - 6      2      7      3      9      3      1      4      0     12     17     29
 6 - 7      1     23      1     24      2      2      1      2      5     51     56
 > 7        1     41      1     36      0     15      0     13      2    105    107
Sunlit      6     71      6     69      6     18      5     15     23    173    196
Shadow      0      0      0      0      2     56      2     59      4    115    119
All         6     71      6     69      8     74      7     74     27    288    315

Latitude   0.0,   solar az. 270.0,  solar el. -30.0,  solar dec.   0.0
           ---------------------------- Quadrant ----------------------------
Magni-     Northwest     Southwest     Southeast     Northeast      All Quadrants
tude       30+    30-    30+    30-    30+    30-    30+    30-    30+    30-    Any
 < 5        0      0      0      0      0      0      0      0      0      0      0
 5 - 6      0      0      0      0      0      0      0      0      0      0      0
 6 - 7      0     11      0      8      0      0      0      0      0     19     19
 > 7        0     21      0     23      0      0      0      0      0     44     44
Sunlit      0     32      0     31      0      0      0      0      0     63     63
Shadow      6     39      6     38      8     74      7     74     27    225    252
All         6     71      6     69      8     74      7     74     27    288    315
```

Appendix C-0-γ. Latitude 0°, solar declination +23.4°

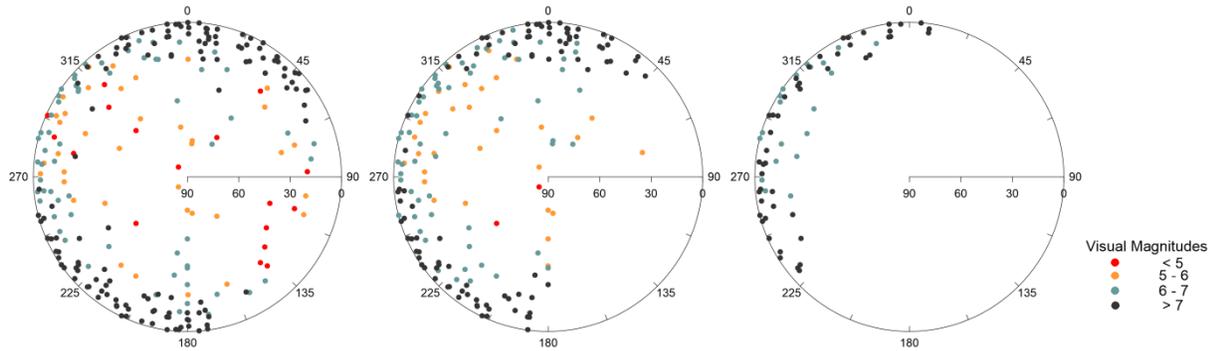

Figure C-0-γ. The distribution and brightness of Starlink satellites is illustrated for an observer at the equator on a date when the Sun is at declination +23.4°. From left to right the Sun's distance below the western horizon is 12° (end of nautical twilight), 18° (end of astronomical twilight) and 30°.

Table C-0-γ. Latitude 0°, solar declination +23.4°

```
Latitude    0.0,   solar az. 294.0,   solar el. -12.0,   solar dec.   23.4
             ---------------------------- Quadrant ----------------------------
Magni-      Northwest      Southwest      Southeast      Northeast      All Quadrants
tude        30+    30-     30+    30-     30+    30-     30+    30-     30+    30-    Any
< 5          2      5       1      0       2      4       1      2       6     11      17
5 - 6        3     14       2      4       3      4       3      4      11     26      37
6 - 7        1     27       2     15       3      6       2      7       8     55      63
> 7          0     25       1     50       0     13       1     43       2    131     133
Sunlit       6     71       6     69       8     27       7     56      27    223     250
Shadow       0      0       0      0       0     47       0     18       0     65      65
All          6     71       6     69       8     74       7     74      27    288     315

Latitude    0.0,   solar az. 294.7,   solar el. -18.0,   solar dec.   23.4
             ---------------------------- Quadrant ----------------------------
Magni-      Northwest      Southwest      Southeast      Northeast      All Quadrants
tude        30+    30-     30+    30-     30+    30-     30+    30-     30+    30-    Any
< 5          0      0       2      0       0      0       0      0       2      0       2
5 - 6        4     16       1      2       4      0       3      0      12     18      30
6 - 7        2     33       2     16       1      0       4      4       9     53      62
> 7          0     22       1     43       0      1       0     28       1     94      95
Sunlit       6     71       6     61       5      1       7     32      24    165     189
Shadow       0      0       0      8       3     73       0     42       3    123     126
All          6     71       6     69       8     74       7     74      27    288     315

Latitude    0.0,   solar az. 297.3,   solar el. -30.0,   solar dec.   23.4
             ---------------------------- Quadrant ----------------------------
Magni-      Northwest      Southwest      Southeast      Northeast      All Quadrants
tude        30+    30-     30+    30-     30+    30-     30+    30-     30+    30-    Any
< 5          0      0       0      0       0      0       0      0       0      0       0
5 - 6        0      0       0      0       0      0       0      0       0      0       0
6 - 7        0     17       0      3       0      0       0      0       0     20      20
> 7          0     29       0     16       0      0       0      3       0     48      48
Sunlit       0     46       0     19       0      0       0      3       0     68      68
Shadow       6     25       6     50       8     74       7     71      27    220     247
All          6     71       6     69       8     74       7     74      27    288     315
```

Appendix C-1-α. Latitude 10°, solar declination -23.4°

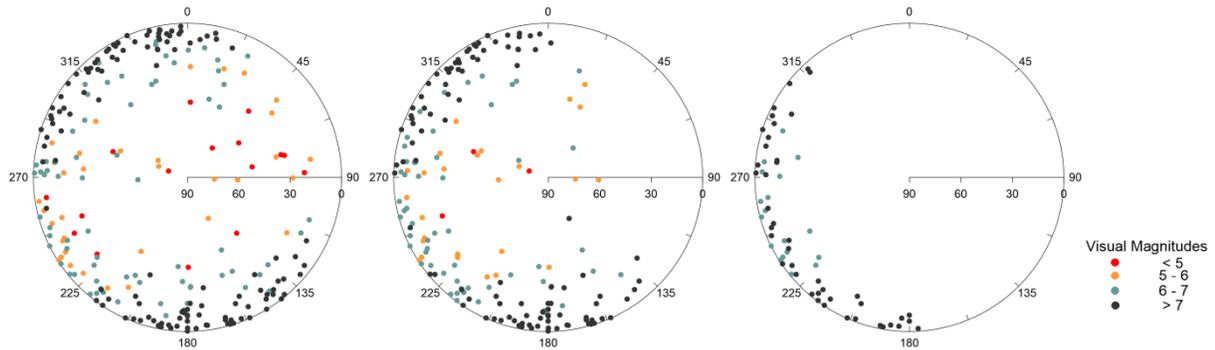

Figure C-1-α. The distribution and brightness of Starlink satellites is illustrated for an observer at latitude +10° at the winter solstice. From left to right the Sun's distance below the western horizon is 12° (end of nautical twilight), 18° (end of astronomical twilight) and 30°.

Table C-1-α. Latitude 10°, solar declination -23.4°

```
Latitude   10.0,   solar az. 248.0,   solar el. -12.0,   solar dec. -23.4
           ---------------------------- Quadrant ----------------------------
Magni-     Northwest     Southwest     Southeast     Northeast     All Quadrants
tude       30+   30-     30+   30-     30+   30-     30+   30-     30+   30-    Any
 < 5        2     0       0     4       2     0       7     1       11    5      16
 5 - 6      3     6       1    13       3     2       1     6        8   27      35
 6 - 7      4    20       2    25       1    11       3     5       10   61      71
 > 7        0    54       0    23       0    40       0     7        0  124     124
Sunlit      9    80       3    65       6    53      11    19       29  217     246
Shadow      0     0       0     0       0    18       0    61        0   79      79
All         9    80       3    65       6    71      11    80       29  296     325

Latitude   10.0,   solar az. 248.5,   solar el. -18.0,   solar dec. -23.4
           ---------------------------- Quadrant ----------------------------
Magni-     Northwest     Southwest     Southeast     Northeast     All Quadrants
tude       30+   30-     30+   30-     30+   30-     30+   30-     30+   30-    Any
 < 5        2     0       0     1       0     0       0     0        2    1       3
 5 - 6      3     5       2    10       3     0       3     0       11   15      26
 6 - 7      4    14       1    30       1     4       1     1        7   49      56
 > 7        0    48       0    24       1    29       0     1        1  102     103
Sunlit      9    67       3    65       5    33       4     2       21  167     188
Shadow      0    13       0     0       1    38       7    78        8  129     137
All         9    80       3    65       6    71      11    80       29  296     325

Latitude   10.0,   solar az. 248.7,   solar el. -30.0,   solar dec. -23.4
           ---------------------------- Quadrant ----------------------------
Magni-     Northwest     Southwest     Southeast     Northeast     All Quadrants
tude       30+   30-     30+   30-     30+   30-     30+   30-     30+   30-    Any
 < 5        0     0       0     0       0     0       0     0        0    0       0
 5 - 6      0     0       0     0       0     0       0     0        0    0       0
 6 - 7      0     5       0    13       0     0       0     0        0   18      18
 > 7        0    15       0    25       0     3       0     0        0   43      43
Sunlit      0    20       0    38       0     3       0     0        0   61      61
Shadow      9    60       3    27       6    68      11    80       29  235     264
All         9    80       3    65       6    71      11    80       29  296     325
```

Appendix C-1-β. Latitude 10°, solar declination 0.0°

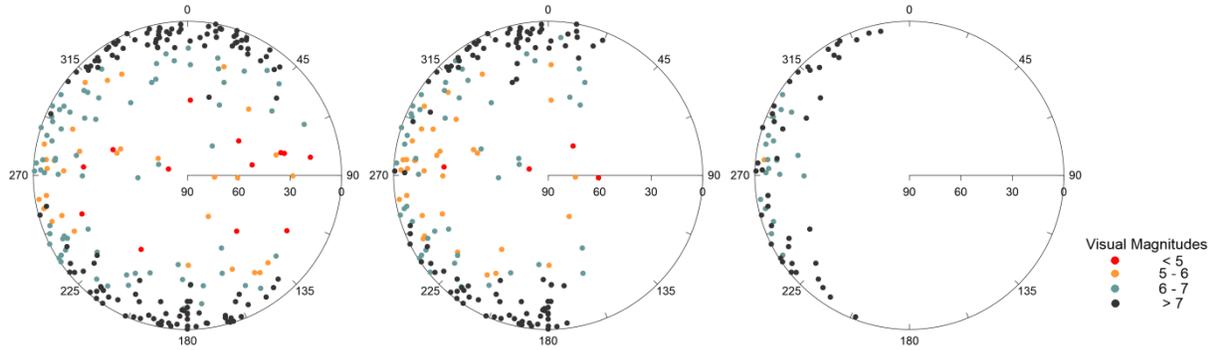

Figure C-1-β. The distribution and brightness of Starlink satellites is illustrated for an observer at latitude +10° at an equinox date. From left to right the Sun's distance below the western horizon is 12° (end of nautical twilight), 18° (end of astronomical twilight) and 30°.

Table C-1-β. Latitude 10°, solar declination 0.0°

```
Latitude   10.0,   solar az. 272.1,   solar el. -12.0,   solar dec.   0.0
                   ------------------------- Quadrant ---------------------------
Magni-     Northwest      Southwest      Southeast      Northeast      All Quadrants
tude       30+    30-     30+    30-     30+    30-     30+    30-     30+    30-    Any
< 5         2      1       1      1       1      1       5      1       9      4      13
5 - 6       3     10       0      6       4      5       2      1       9     22      31
6 - 7       4     29       2     21       1      7       3      9      10     66      76
> 7         0     40       0     37       0     26       1     29       1    132     133
Sunlit      9     80       3     65       6     39      11     40      29    224     253
Shadow      0      0       0      0       0     32       0     40       0     72      72
All         9     80       3     65       6     71      11     80      29    296     325

Latitude   10.0,   solar az. 273.3,   solar el. -18.0,   solar dec.   0.0
                   ------------------------- Quadrant ---------------------------
Magni-     Northwest      Southwest      Southeast      Northeast      All Quadrants
tude       30+    30-     30+    30-     30+    30-     30+    30-     30+    30-    Any
< 5         1      1       0      0       1      0       1      0       3      1       4
5 - 6       2     16       1     10       3      0       1      1       7     27      34
6 - 7       4     20       2     16       1      2       3      2      10     40      50
> 7         2     43       0     39       0     12       0     11       2    105     107
Sunlit      9     80       3     65       5     14       5     14      22    173     195
Shadow      0      0       0      0       1     57       6     66       7    123     130
All         9     80       3     65       6     71      11     80      29    296     325

Latitude   10.0,   solar az. 275.8,   solar el. -30.0,   solar dec.   0.0
                   ------------------------- Quadrant ---------------------------
Magni-     Northwest      Southwest      Southeast      Northeast      All Quadrants
tude       30+    30-     30+    30-     30+    30-     30+    30-     30+    30-    Any
< 5         0      0       0      0       0      0       0      0       0      0       0
5 - 6       0      1       0      0       0      0       0      0       0      1       1
6 - 7       0     13       0      6       0      0       0      0       0     19      19
> 7         0     24       0     18       0      0       0      0       0     42      42
Sunlit      0     38       0     24       0      0       0      0       0     62      62
Shadow      9     42       3     41       6     71      11     80      29    234     263
All         9     80       3     65       6     71      11     80      29    296     325
```

Appendix C-1-γ. Latitude 10°, solar declination +23.4°

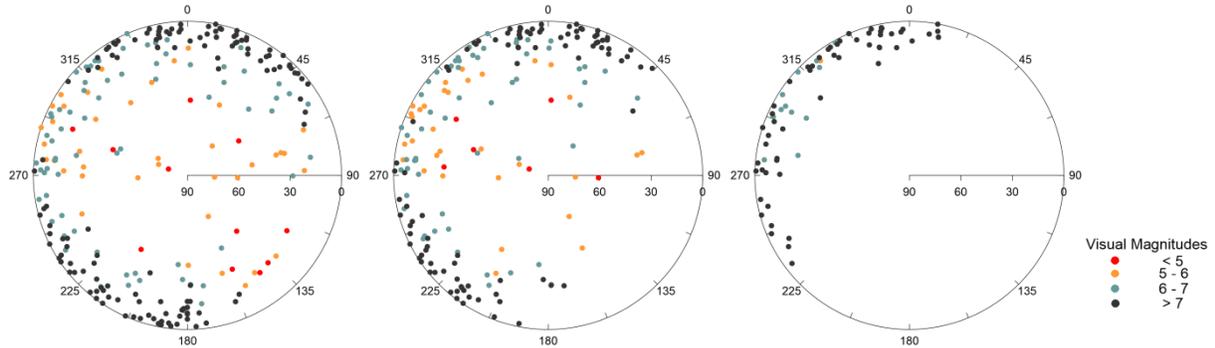

Figure C-1-γ. The distribution and brightness of Starlink satellites is illustrated for an observer at latitude +10° at the summer solstice. From left to right the Sun's distance below the western horizon is 12° (end of nautical twilight), 18° (end of astronomical twilight) and 30°.

Table C-1-γ. Latitude 10°, solar declination +23.4°

```
Latitude   10.0,   solar az. 296.7,   solar el. -12.0,   solar dec.   23.4
          ---------------------------- Quadrant ----------------------------
Magni-    Northwest     Southwest     Southeast     Northeast     All Quadrants
tude      30+    30-    30+    30-    30+    30-    30+    30-    30+    30-    Any
 < 5       2      1      1      0      1      4      2      0      6      5     11
 5 - 6     4     15      1      2      4      4      6      3     15     24     39
 6 - 7     3     33      1     15      1      4      3     10      8     62     70
 > 7       0     31      0     48      0     12      0     42      0    133    133
Sunlit     9     80      3     65      6     24     11     55     29    224    253
Shadow     0      0      0      0      0     47      0     25      0     72     72
All        9     80      3     65      6     71     11     80     29    296    325

Latitude   10.0,   solar az. 298.8,   solar el. -18.0,   solar dec.   23.4
          ---------------------------- Quadrant ----------------------------
Magni-    Northwest     Southwest     Southeast     Northeast     All Quadrants
tude      30+    30-    30+    30-    30+    30-    30+    30-    30+    30-    Any
 < 5       2      2      0      0      1      0      1      0      4      2      6
 5 - 6     2     19      2      2      3      0      3      1     10     22     32
 6 - 7     5     30      1     11      0      1      3      6      9     48     57
 > 7       0     29      0     37      0      2      0     29      0     97     97
Sunlit     9     80      3     50      4      3      7     36     23    169    192
Shadow     0      0      0     15      2     68      4     44      6    127    133
All        9     80      3     65      6     71     11     80     29    296    325

Latitude   10.0,   solar az. 304.6,   solar el. -30.0,   solar dec.   23.4
          ---------------------------- Quadrant ----------------------------
Magni-    Northwest     Southwest     Southeast     Northeast     All Quadrants
tude      30+    30-    30+    30-    30+    30-    30+    30-    30+    30-    Any
 < 5       0      0      0      0      0      0      0      0      0      0      0
 5 - 6     0      1      0      0      0      0      0      0      0      1      1
 6 - 7     0     12      0      0      0      0      0      0      0     12     12
 > 7       0     42      0      8      0      0      0      6      0     56     56
Sunlit     0     55      0      8      0      0      0      6      0     69     69
Shadow     9     25      3     57      6     71     11     74     29    227    256
All        9     80      3     65      6     71     11     80     29    296    325
```

Appendix C-2-α. Latitude 20°, solar declination -23.4°

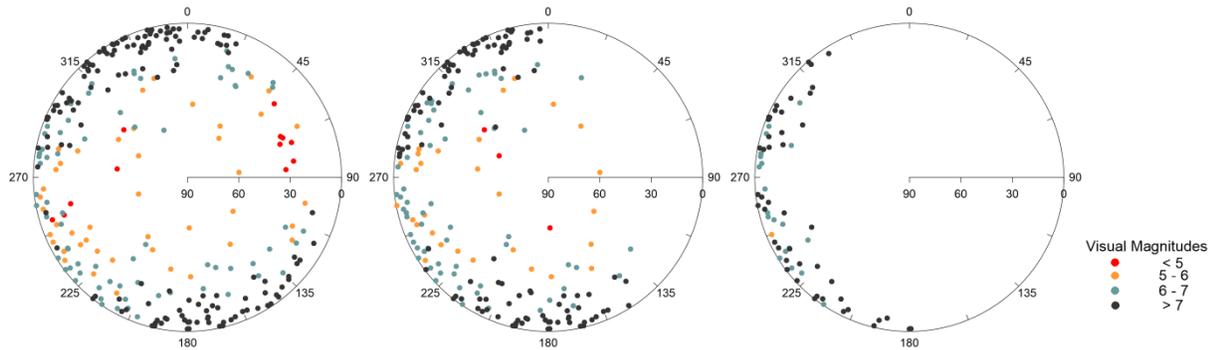

Figure C-2-α. The distribution and brightness of Starlink satellites is illustrated for an observer at latitude +20° at the winter solstice. From left to right the Sun's distance below the western horizon is 12° (end of nautical twilight), 18° (end of astronomical twilight) and 30°.

Table C-2-α. Latitude 20°, solar declination -23.4°

```
Latitude   20.0,   solar az. 249.2,   solar el. -12.0,   solar dec. -23.4
           ---------------------------- Quadrant ----------------------------
Magni-     Northwest    Southwest    Southeast    Northeast    All Quadrants
tude       30+   30-    30+   30-    30+   30-    30+   30-    30+   30-   Any
 < 5        2     0      0     3      0     0      3     4      5     7    12
 5 - 6      4     5      4    16      4     3      5     3     17    27    44
 6 - 7      3    18      2    27      3     9      0     9      8    63    71
 > 7        0    73      0    20      0    40      0    12      0   145   145
Sunlit      9    96      6    66      7    52      8    28     30   242   272
Shadow      0     0      0     0      0    21      0    63      0    84    84
All         9    96      6    66      7    73      8    91     30   326   356

Latitude   20.0,   solar az. 251.0,   solar el. -18.0,   solar dec. -23.4
           ---------------------------- Quadrant ----------------------------
Magni-     Northwest    Southwest    Southeast    Northeast    All Quadrants
tude       30+   30-    30+   30-    30+   30-    30+   30-    30+   30-   Any
 < 5        2     0      0     0      1     0      0     0      3     0     3
 5 - 6      3     7      3     9      4     0      3     0     13    16    29
 6 - 7      3    19      2    30      2     5      0     1      7    55    62
 > 7        1    62      1    27      0    25      0     0      2   114   116
Sunlit      9    88      6    66      7    30      3     1     25   185   210
Shadow      0     8      0     0      0    43      5    90      5   141   146
All         9    96      6    66      7    73      8    91     30   326   356

Latitude   20.0,   solar az. 253.9,   solar el. -30.0,   solar dec. -23.4
           ---------------------------- Quadrant ----------------------------
Magni-     Northwest    Southwest    Southeast    Northeast    All Quadrants
tude       30+   30-    30+   30-    30+   30-    30+   30-    30+   30-   Any
 < 5        0     0      0     0      0     0      0     0      0     0     0
 5 - 6      0     0      0     1      0     0      0     0      0     1     1
 6 - 7      0     8      0    12      0     0      0     0      0    20    20
 > 7        0    19      0    25      0     2      0     0      0    46    46
Sunlit      0    27      0    38      0     2      0     0      0    67    67
Shadow      9    69      6    28      7    71      8    91     30   259   289
All         9    96      6    66      7    73      8    91     30   326   356
```

Appendix C-2-β. Latitude 20°, solar declination 0.0°

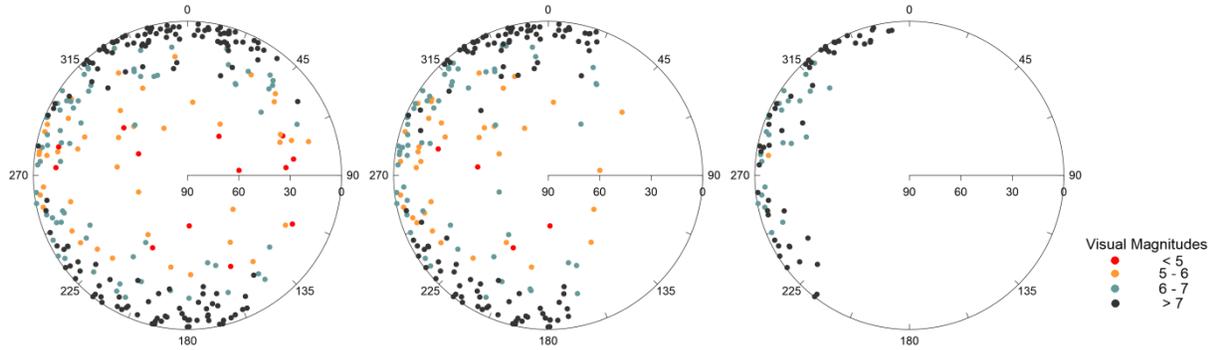

Figure C-2-β. The distribution and brightness of Starlink satellites is illustrated for an observer at latitude +20° at an equinox date. From left to right the Sun's distance below the western horizon is 12° (end of nautical twilight), 18° (end of astronomical twilight) and 30°.

Table C-2-β. Latitude 20°, solar declination 0.0°

```
Latitude   20.0,   solar az. 274.4,   solar el. -12.0,   solar dec.   0.0
                     ---------------------------- Quadrant ----------------------------
Magni-     Northwest      Southwest      Southeast      Northeast      All Quadrants
tude       30+    30-     30+    30-    30+    30-     30+    30-     30+    30-    Any
 < 5        2      2       1      0      2      1       3      2       8      5      13
 5 - 6      6     12       3      8      3      2       4      5      16     27      43
 6 - 7      1     32       2     20      2      5       1     10       6     67      73
 > 7        0     50       0     38      0     25       0     33       0    146     146
Sunlit      9     96       6     66      7     33       8     50      30    245     275
Shadow      0      0       0      0      0     40       0     41       0     81      81
All         9     96       6     66      7     73       8     91      30    326     356

Latitude   20.0,   solar az. 276.8,   solar el. -18.0,   solar dec.   0.0
                     ---------------------------- Quadrant ----------------------------
Magni-     Northwest      Southwest      Southeast      Northeast      All Quadrants
tude       30+    30-     30+    30-    30+    30-     30+    30-     30+    30-    Any
 < 5        1      1       1      0      1      0       0      0       3      1       4
 5 - 6      6     14       2      9      2      0       3      0      13     23      36
 6 - 7      1     36       3     15      2      1       1      3       7     55      62
 > 7        1     45       0     42      0     10       0     16       1    113     114
Sunlit      9     96       6     66      5     11       4     19      24    192     216
Shadow      0      0       0      0      2     62       4     72       6    134     140
All         9     96       6     66      7     73       8     91      30    326     356

Latitude   20.0,   solar az. 282.1,   solar el. -30.0,   solar dec.   0.0
                     ---------------------------- Quadrant ----------------------------
Magni-     Northwest      Southwest      Southeast      Northeast      All Quadrants
tude       30+    30-     30+    30-    30+    30-     30+    30-     30+    30-    Any
 < 5        0      0       0      0      0      0       0      0       0      0       0
 5 - 6      0      1       0      0      0      0       0      0       0      1       1
 6 - 7      0     18       0      6      0      0       0      0       0     24      24
 > 7        0     35       0     16      0      0       0      0       0     51      51
Sunlit      0     54       0     22      0      0       0      0       0     76      76
Shadow      9     42       6     44      7     73       8     91      30    250     280
All         9     96       6     66      7     73       8     91      30    326     356
```

Appendix C-2-γ. Latitude 20°, solar declination +23.4°

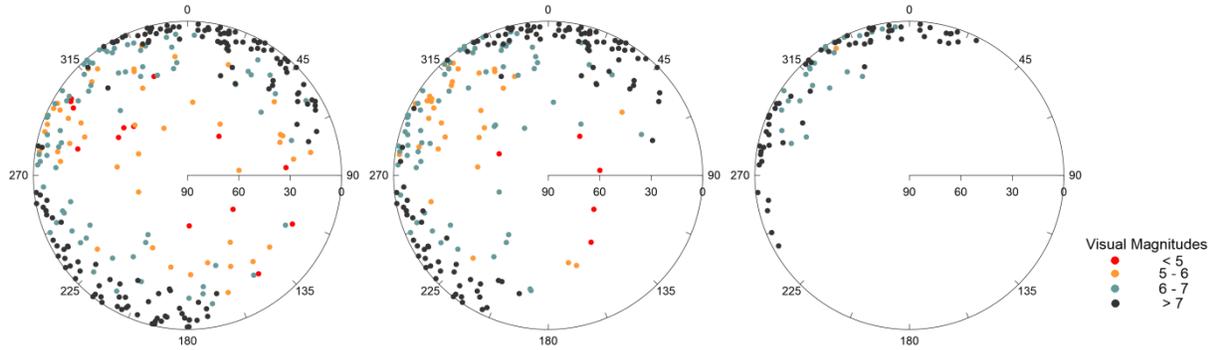

Figure C-2-γ. The distribution and brightness of Starlink satellites is illustrated for an observer at latitude +20° at the summer solstice. From left to right the Sun's distance below the western horizon is 12° (end of nautical twilight), 18° (end of astronomical twilight) and 30°.

Table C-2-γ. Latitude 20°, solar declination +23.4°

```
Latitude   20.0,   solar az. 300.6,   solar el. -12.0,   solar dec.   23.4
         ---------------------------- Quadrant ----------------------------
Magni-   Northwest      Southwest      Southeast      Northeast     All Quadrants
tude     30+    30-     30+    30-     30+    30-     30+    30-    30+    30-    Any
 < 5      3      5       0      0       2      2       2      0      7      7     14
 5 - 6    5     20       3      1       4      3       5      5     17     29     46
 6 - 7    1     43       3     14       1      6       1      9      6     72     78
 > 7      0     28       0     51       0      9       0     58      0    146    146
Sunlit    9     96       6     66       7     20       8     72     30    254    284
Shadow    0      0       0      0       0     53       0     19      0     72     72
All       9     96       6     66       7     73       8     91     30    326    356

Latitude   20.0,   solar az. 304.2,   solar el. -18.0,   solar dec.   23.4
         ---------------------------- Quadrant ----------------------------
Magni-   Northwest      Southwest      Southeast      Northeast     All Quadrants
tude     30+    30-     30+    30-     30+    30-     30+    30-    30+    30-    Any
 < 5      1      0       0      0       2      0       2      0      5      0      5
 5 - 6    5     21       0      2       2      0       1      0      8     23     31
 6 - 7    3     48       4     10       0      0       3      3     10     61     71
 > 7      0     27       0     39       0      0       0     46      0    112    112
Sunlit    9     96       4     51       4      0       6     49     23    196    219
Shadow    0      0       2     15       3     73       2     42      7    130    137
All       9     96       6     66       7     73       8     91     30    326    356

Latitude   20.0,   solar az. 314.3,   solar el. -30.0,   solar dec.   23.4
         ---------------------------- Quadrant ----------------------------
Magni-   Northwest      Southwest      Southeast      Northeast     All Quadrants
tude     30+    30-     30+    30-     30+    30-     30+    30-    30+    30-    Any
 < 5      0      0       0      0       0      0       0      0      0      0      0
 5 - 6    0      1       0      0       0      0       0      0      0      1      1
 6 - 7    0     25       0      0       0      0       0      0      0     25     25
 > 7      0     43       0      5       0      0       0     12      0     60     60
Sunlit    0     69       0      5       0      0       0     12      0     86     86
Shadow    9     27       6     61       7     73       8     79     30    240    270
All       9     96       6     66       7     73       8     91     30    326    356
```

## Appendix C-3-α. Latitude 30°, solar declination -23.4°

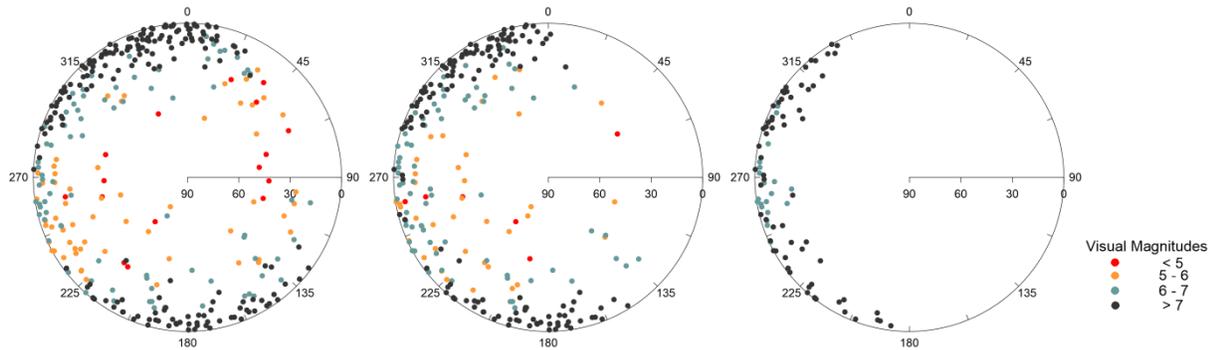

Figure C-3-α. The distribution and brightness of Starlink satellites is illustrated for an observer at latitude +30° at the winter solstice. From left to right the Sun's distance below the western horizon is 12° (end of nautical twilight), 18° (end of astronomical twilight) and 30°.

## Table C-3-α. Latitude 30°, solar declination -23.4°

```
Latitude   30.0,   solar az. 249.8,   solar el. -12.0,   solar dec. -23.4
          ----------------------------- Quadrant ----------------------------
Magni-    Northwest      Southwest      Southeast      Northeast       All Quadrants
tude      30+    30-     30+    30-     30+    30-     30+    30-     30+    30-     Any
 < 5       2      0       3      3       2      0       3      3      10      6       16
5 - 6      2      6       7     25       3      6       5      4      17     41       58
6 - 7      4     22       1     29       3     12       1      9       9     72       81
 > 7       0     98       0     28       0     35       0     26       0    187      187
Sunlit     8    126      11     85       8     53       9     42      36    306      342
Shadow     0      0       0      0       0     29       0     88       0    117      117
All        8    126      11     85       8     82       9    130      36    423      459

Latitude   30.0,   solar az. 252.9,   solar el. -18.0,   solar dec. -23.4
          ----------------------------- Quadrant ----------------------------
Magni-    Northwest      Southwest      Southeast      Northeast       All Quadrants
tude      30+    30-     30+    30-     30+    30-     30+    30-     30+    30-     Any
 < 5       0      0       3      2       0      0       1      0       4      2        6
5 - 6      4      6       7     18       2      0       1      0      14     24       38
6 - 7      4     26       1     30       3      4       1      0       9     60       69
 > 7       0     84       0     35       0     22       0      2       0    143      143
Sunlit     8    116      11     85       5     26       3      2      27    229      256
Shadow     0     10       0      0       3     56       6    128       9    194      203
All        8    126      11     85       8     82       9    130      36    423      459

Latitude   30.0,   solar az. 258.7,   solar el. -30.0,   solar dec. -23.4
          ----------------------------- Quadrant ----------------------------
Magni-    Northwest      Southwest      Southeast      Northeast       All Quadrants
tude      30+    30-     30+    30-     30+    30-     30+    30-     30+    30-     Any
 < 5       0      0       0      0       0      0       0      0       0      0        0
5 - 6      0      0       0      0       0      0       0      0       0      0        0
6 - 7      0      6       0     14       0      0       0      0       0     20       20
 > 7       0     34       0     31       0      0       0      0       0     65       65
Sunlit     0     40       0     45       0      0       0      0       0     85       85
Shadow     8     86      11     40       8     82       9    130      36    338      374
All        8    126      11     85       8     82       9    130      36    423      459
```

Appendix C-3-β. Latitude 30°, solar declination 0.0°

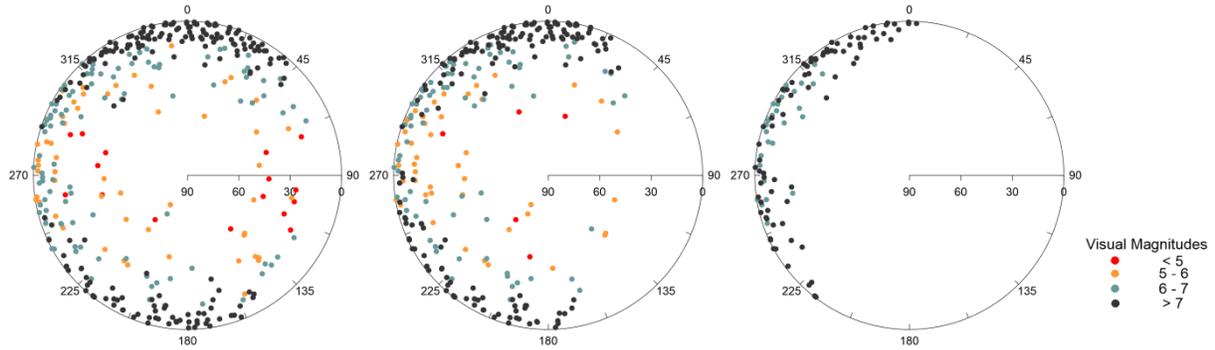

Figure C-3-β. The distribution and brightness of Starlink satellites is illustrated for an observer at latitude +30° at an equinox date. From left to right the Sun's distance below the western horizon is 12° (end of nautical twilight), 18° (end of astronomical twilight) and 30°.

Table C-3-β. Latitude 30°, solar declination 0.0°

```
Latitude   30.0,   solar az. 277.0,   solar el. -12.0,   solar dec.    0.0
                -------------------------- Quadrant ---------------------------
Magni-    Northwest      Southwest      Southeast      Northeast      All Quadrants
tude      30+    30-     30+    30-     30+    30-     30+    30-     30+    30-    Any
 < 5       2      2       2      1       3      4       1      1       8      8      16
 5 - 6     2     19       7      9       4      5       5      2      18     35      53
 6 - 7     2     40       2     33       1      7       3     13       8     93     101
 > 7       2     65       0     42       0     23       0     63       2    193     195
Sunlit     8    126      11     85       8     39       9     79      36    329     365
Shadow     0      0       0      0       0     43       0     51       0     94      94
All        8    126      11     85       8     82       9    130      36    423     459

Latitude   30.0,   solar az. 280.8,   solar el. -18.0,   solar dec.    0.0
                -------------------------- Quadrant ---------------------------
Magni-    Northwest      Southwest      Southeast      Northeast      All Quadrants
tude      30+    30-     30+    30-     30+    30-     30+    30-     30+    30-    Any
 < 5       1      1       2      0       0      0       1      0       4      1       5
 5 - 6     3     18       5     11       4      0       3      0      15     29      44
 6 - 7     4     38       4     19       1      2       1      6      10     65      75
 > 7       0     69       0     53       0      7       0     38       0    167     167
Sunlit     8    126      11     83       5      9       5     44      29    262     291
Shadow     0      0       0      2       3     73       4     86       7    161     168
All        8    126      11     85       8     82       9    130      36    423     459

Latitude   30.0,   solar az. 289.5,   solar el. -30.0,   solar dec.    0.0
                -------------------------- Quadrant ---------------------------
Magni-    Northwest      Southwest      Southeast      Northeast      All Quadrants
tude      30+    30-     30+    30-     30+    30-     30+    30-     30+    30-    Any
 < 5       0      0       0      0       0      0       0      0       0      0       0
 5 - 6     0      0       0      0       0      0       0      0       0      0       0
 6 - 7     0     22       0      6       0      0       0      0       0     28      28
 > 7       0     55       0     25       0      0       0      1       0     81      81
Sunlit     0     77       0     31       0      0       0      1       0    109     109
Shadow     8     49      11     54       8     82       9    129      36    314     350
All        8    126      11     85       8     82       9    130      36    423     459
```

Appendix C-3-γ. Latitude 30°, solar declination +23.4°

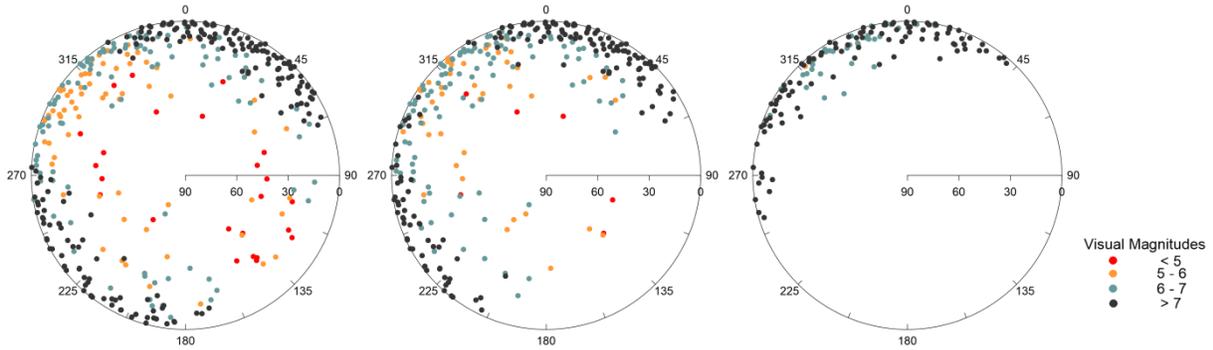

Figure C-3-γ. The distribution and brightness of Starlink satellites is illustrated for an observer at latitude +30° at the summer solstice. From left to right the Sun's distance below the western and northwestern horizon is 12° (end of nautical twilight), 18° (end of astronomical twilight) and 30°.

Table C-3-γ. Latitude 30°, solar declination +23.4°

```
Latitude   30.0,   solar az. 306.3,   solar el. -12.0,   solar dec.   23.4
          ---------------------------- Quadrant ----------------------------
Magni-    Northwest      Southwest      Southeast      Northeast      All Quadrants
tude      30+    30-     30+    30-    30+    30-     30+    30-     30+    30-     Any
 < 5       3      3       3      0      5      6       4      0       15     9       24
 5 - 6     3     39       6      5      2      5       2      2       13    51       64
 6 - 7     2     55       1     16      1      6       3     16        7    93      100
 > 7       0     29       1     63      0      3       0     93        1   188      189
Sunlit     8    126      11     84      8     20       9    111       36   341      377
Shadow     0      0       0      1      0     62       0     19        0    82       82
All        8    126      11     85      8     82       9    130       36   423      459

Latitude   30.0,   solar az. 312.1,   solar el. -18.0,   solar dec.   23.4
          ---------------------------- Quadrant ----------------------------
Magni-    Northwest      Southwest      Southeast      Northeast      All Quadrants
tude      30+    30-     30+    30-    30+    30-     30+    30-     30+    30-     Any
 < 5       1      1       1      0      2      0       1      0        5     1        6
 5 - 6     5     25       4      1      3      0       1      2       13    28       41
 6 - 7     2     64       5     12      0      0       5     12       12    88      100
 > 7       0     36       0     44      0      0       0     79        0   159      159
Sunlit     8    126      10     57      5      0       7     93       30   276      306
Shadow     0      0       1     28      3     82       2     37        6   147      153
All        8    126      11     85      8     82       9    130       36   423      459

Latitude   30.0,   solar az. 329.6,   solar el. -30.0,   solar dec.   23.4
          ---------------------------- Quadrant ----------------------------
Magni-    Northwest      Southwest      Southeast      Northeast      All Quadrants
tude      30+    30-     30+    30-    30+    30-     30+    30-     30+    30-     Any
 < 5       0      0       0      0      0      0       0      0        0     0        0
 5 - 6     0      1       0      0      0      0       0      0        0     1        1
 6 - 7     0     30       0      0      0      0       0      2        0    32       32
 > 7       0     63       0      7      0      0       0     33        0   103      103
Sunlit     0     94       0      7      0      0       0     35        0   136      136
Shadow     8     32      11     78      8     82       9     95       36   287      323
All        8    126      11     85      8     82       9    130       36   423      459
```

Appendix C-4-α. Latitude 40°, solar declination -23.4°

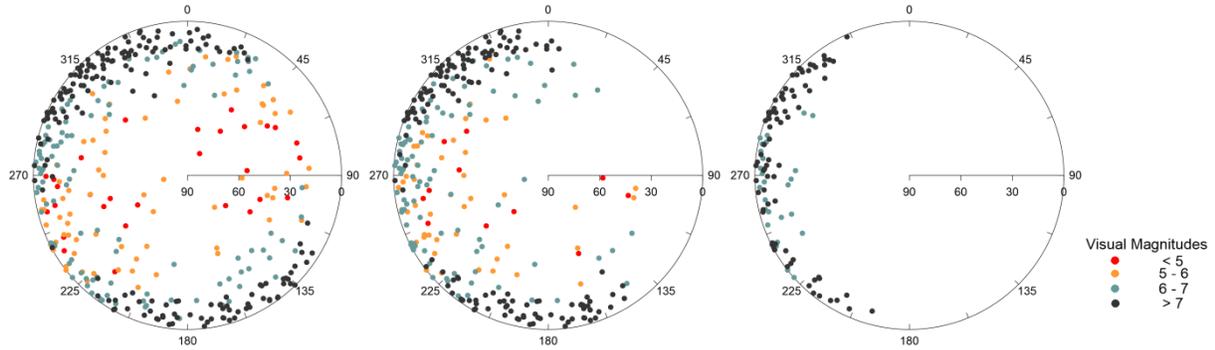

Figure C-4-α. The distribution and brightness of Starlink satellites is illustrated for an observer at latitude +40° at the winter solstice. From left to right the Sun's distance below the western horizon is 12° (end of nautical twilight), 18° (end of astronomical twilight) and 30°.

Table C-4-α. Latitude 40°, solar declination -23.4°

```
Latitude   40.0,   solar az. 249.4,   solar el. -12.0,   solar dec. -23.4
                        ---------------------------- Quadrant ----------------------------
Magni-     Northwest       Southwest       Southeast       Northeast       All Quadrants
tude       30+    30-      30+    30-     30+    30-     30+    30-      30+    30-   Any
 < 5        1      1        5      8       4      0       8      2       18     11     29
 5 - 6      7      5        8     27       8      2       4     11       27     45     72
 6 - 7      2     38        2     34       3     15       1     12        8     99    107
 > 7        2     89        0     29       0     42       0     14        2    174    176
Sunlit     12    133       15     98      15     59      13     39       55    329    384
Shadow      0      0        0      0       0     34       0     98        0    132    132
All        12    133       15     98      15     93      13    137       55    461    516

Latitude   40.0,   solar az. 254.2,   solar el. -18.0,   solar dec. -23.4
                        ---------------------------- Quadrant ----------------------------
Magni-     Northwest       Southwest       Southeast       Northeast       All Quadrants
tude       30+    30-      30+    30-     30+    30-     30+    30-      30+    30-   Any
 < 5        2      1        2      3       3      0       0      0        7      4     11
 5 - 6      5      8        7     19       4      1       0      0       16     28     44
 6 - 7      5     37        6     35       3      2       2      1       16     75     91
 > 7        0     82        0     41       0     24       0      3        0    150    150
Sunlit     12    128       15     98      10     27       2      4       39    257    296
Shadow      0      5        0      0       5     66      11    133       16    204    220
All        12    133       15     98      15     93      13    137       55    461    516

Latitude   40.0,   solar az. 263.4,   solar el. -30.0,   solar dec. -23.4
                        ---------------------------- Quadrant ----------------------------
Magni-     Northwest       Southwest       Southeast       Northeast       All Quadrants
tude       30+    30-      30+    30-     30+    30-     30+    30-      30+    30-   Any
 < 5        0      0        0      0       0      0       0      0        0      0      0
 5 - 6      0      0        0      0       0      0       0      0        0      0      0
 6 - 7      0     11        0     15       0      0       0      0        0     26     26
 > 7        0     48        0     29       0      0       0      0        0     77     77
Sunlit      0     59        0     44       0      0       0      0        0    103    103
Shadow     12     74       15     54      15     93      13    137       55    358    413
All        12    133       15     98      15     93      13    137       55    461    516
```

Appendix C-4-β. Latitude 40°, solar declination 0.0°

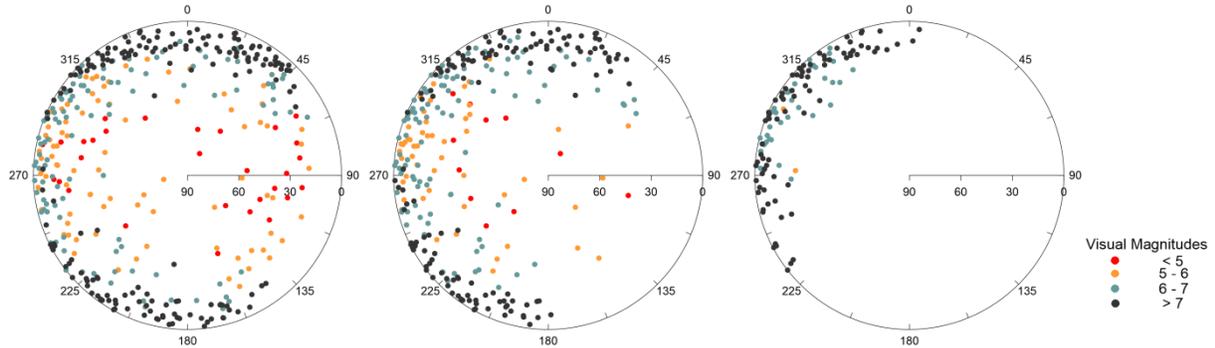

Figure C-4-β. The distribution and brightness of Starlink satellites is illustrated for an observer at latitude +40° at an equinox date. From left to right the Sun's distance below the western horizon is 12° (end of nautical twilight), 18° (end of astronomical twilight) and 30°.

Table C-4-β. Latitude 40°, solar declination 0.0°

```
Latitude   40.0,   solar az. 280.3,   solar el. -12.0,   solar dec.    0.0
                --------------------------- Quadrant ---------------------------
Magni-     Northwest      Southwest      Southeast      Northeast      All Quadrants
tude       30+    30-     30+    30-    30+    30-     30+    30-     30+    30-    Any
 < 5        4      4       1      3      7      1       6      3      18     11     29
 5 - 6      4     31      10     12      8      7       5      6      27     56     83
 6 - 7      3     45       3     29      0      7       2     17       8     98    106
 > 7        1     53       1     54      0     17       0     49       2    173    175
Sunlit     12    133      15     98     15     32      13     75      55    338    393
Shadow      0      0       0      0      0     61       0     62       0    123    123
All        12    133      15     98     15     93      13    137      55    461    516

Latitude   40.0,   solar az. 285.8,   solar el. -18.0,   solar dec.    0.0
                --------------------------- Quadrant ---------------------------
Magni-     Northwest      Southwest      Southeast      Northeast      All Quadrants
tude       30+    30-     30+    30-    30+    30-     30+    30-     30+    30-    Any
 < 5        4      2       3      0      1      0       1      0       9      2     11
 5 - 6      3     27       6      8      4      0       2      0      15     35     50
 6 - 7      5     55       6     24      0      0       1      8      12     87     99
 > 7        0     49       0     59      0      1       1     30       1    139    140
Sunlit     12    133      15     91      5      1       5     38      37    263    300
Shadow      0      0       0      7     10     92       8     99      18    198    216
All        12    133      15     98     15     93      13    137      55    461    516

Latitude   40.0,   solar az. 299.0,   solar el. -30.0,   solar dec.    0.0
                --------------------------- Quadrant ---------------------------
Magni-     Northwest      Southwest      Southeast      Northeast      All Quadrants
tude       30+    30-     30+    30-    30+    30-     30+    30-     30+    30-    Any
 < 5        0      0       0      0      0      0       0      0       0      0      0
 5 - 6      0      2       0      0      0      0       0      0       0      2      2
 6 - 7      0     34       0      1      0      0       0      0       0     35     35
 > 7        0     59       0     18      0      0       0      2       0     79     79
Sunlit      0     95       0     19      0      0       0      2       0    116    116
Shadow     12     38      15     79     15     93      13    135      55    345    400
All        12    133      15     98     15     93      13    137      55    461    516
```

Appendix C-4-γ. Latitude 40°, solar declination +23.4°

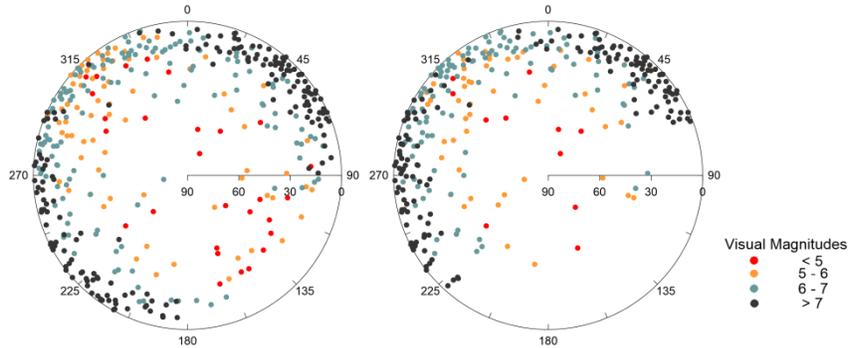

Figure C-4-γ. The distribution and brightness of Starlink satellites is illustrated for an observer at latitude +40° at the summer solstice. From left to right the Sun's distance below the northwestern horizon is 12° (end of nautical twilight) and 18° (end of astronomical twilight). The Sun does not descend to 30° below the horizon at summer solstice at this latitude.

Table C-4-γ. Latitude 40°, solar declination +23.4°

```
Latitude   40.0,   solar az. 315.1,   solar el. -12.0,   solar dec.   23.4
           ----------------------------- Quadrant ----------------------------
Magni-     Northwest      Southwest      Southeast      Northeast      All Quadrants
tude       30+    30-     30+    30-     30+    30-     30+    30-     30+    30-    Any
 < 5         3      6       2      0       8      4       4      1      17     11     28
 5 - 6       6     36       7      3       6      5       5      4      24     48     72
 6 - 7       3     70       5     22       1      6       4     28      13    126    139
 > 7         0     21       1     65       0      0       0     86       1    172    173
Sunlit      12    133      15     90      15     15      13    119      55    357    412
Shadow       0      0       0      8       0     78       0     18       0    104    104
All         12    133      15     98      15     93      13    137      55    461    516

Latitude   40.0,   solar az. 324.9,   solar el. -18.0,   solar dec.   23.4
           ----------------------------- Quadrant ----------------------------
Magni-     Northwest      Southwest      Southeast      Northeast      All Quadrants
tude       30+    30-     30+    30-     30+    30-     30+    30-     30+    30-    Any
 < 5         2      3       1      0       2      0       3      0       8      3     11
 5 - 6       8     22       8      0       3      0       4      4      23     26     49
 6 - 7       2     69       4     11       1      0       4     18      11     98    109
 > 7         0     39       0     40       0      0       0     74       0    153    153
Sunlit      12    133      13     51       6      0      11     96      42    280    322
Shadow       0      0       2     47       9     93       2     41      13    181    194
All         12    133      15     98      15     93      13    137      55    461    516
```

Appendix C-5-α. Latitude 50°, solar declination -23.4°

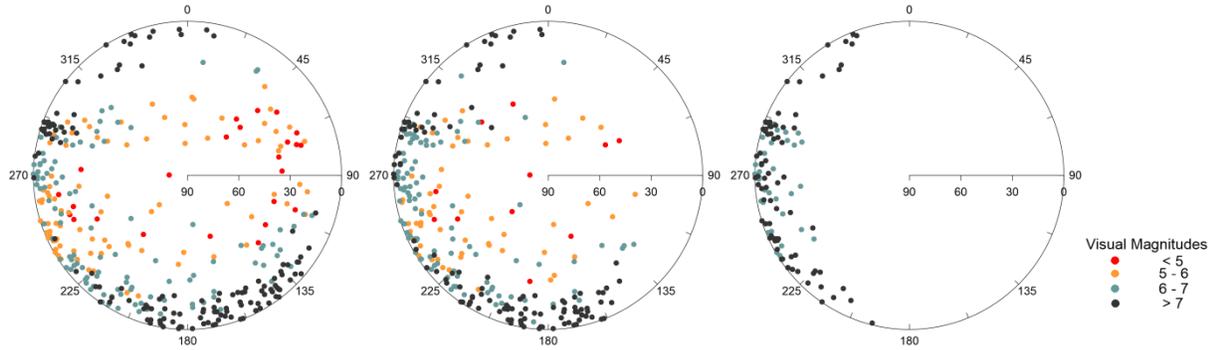

Figure C-5-α. The distribution and brightness of Starlink satellites is illustrated for an observer at latitude +50° at the winter solstice. From left to right the Sun's distance below the western horizon is 12° (end of nautical twilight), 18° (end of astronomical twilight) and 30°.

Table C-5-α. Latitude 50°, solar declination -23.4°

```
Latitude   50.0,   solar az. 247.8,   solar el. -12.0,   solar dec. -23.4
                       ---------------------- Quadrant ----------------------
Magni-     Northwest     Southwest     Southeast     Northeast     All Quadrants
tude       30+    30-    30+    30-   30+    30-    30+    30-    30+    30-    Any
 < 5        1      1      2      4     4      1      6      5     13     11     24
 5 - 6      9      7      9     41     5      4     10      4     33     56     89
 6 - 7      7     17      3     56     4     20      0      3     14     96    110
 > 7        0     38      0     32     0     60      0      2      0    132    132
Sunlit     17     63     14    133    13     85     16     14     60    295    355
Shadow      0      0      0      0     0     51      0     51      0    102    102
All        17     63     14    133    13    136     16     65     60    397    457

Latitude   50.0,   solar az. 254.8,   solar el. -18.0,   solar dec. -23.4
                       ---------------------- Quadrant ----------------------
Magni-     Northwest     Southwest     Southeast     Northeast     All Quadrants
tude       30+    30-    30+    30-   30+    30-    30+    30-    30+    30-    Any
 < 5        3      0      2      3     1      0      2      0      8      3     11
 5 - 6      7      5      8     26     6      1      4      0     25     32     57
 6 - 7      5     23      4     57     2      6      0      1     11     87     98
 > 7        2     35      0     47     0     32      0      0      2    114    116
Sunlit     17     63     14    133     9     39      6      1     46    236    282
Shadow      0      0      0      0     4     97     10     64     14    161    175
All        17     63     14    133    13    136     16     65     60    397    457

Latitude   50.0,   solar az. 268.5,   solar el. -30.0,   solar dec. -23.4
                       ---------------------- Quadrant ----------------------
Magni-     Northwest     Southwest     Southeast     Northeast     All Quadrants
tude       30+    30-    30+    30-   30+    30-    30+    30-    30+    30-    Any
 < 5        0      0      0      0     0      0      0      0      0      0      0
 5 - 6      0      0      0      0     0      0      0      0      0      0      0
 6 - 7      0     14      0     17     0      0      0      0      0     31     31
 > 7        0     34      0     42     0      0      0      0      0     76     76
Sunlit      0     48      0     59     0      0      0      0      0    107    107
Shadow     17     15     14     74    13    136     16     65     60    290    350
All        17     63     14    133    13    136     16     65     60    397    457
```

Appendix C-5-β. Latitude 50°, solar declination 0.0°

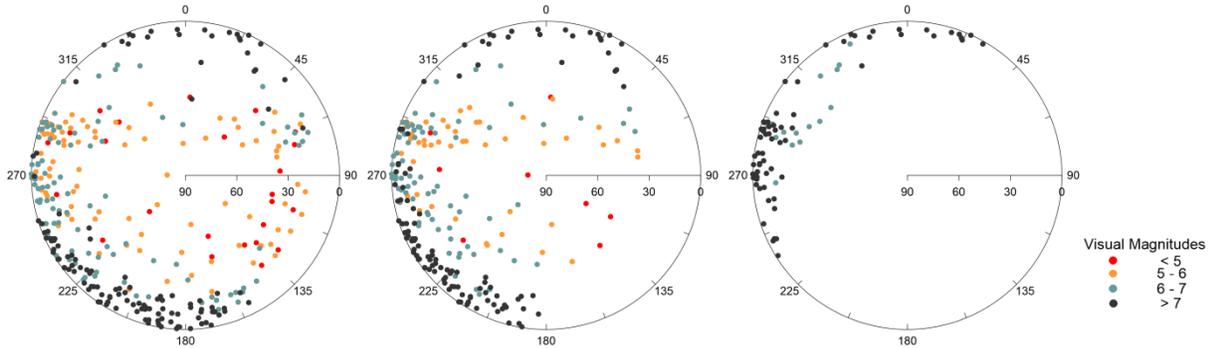

Figure C-5-β. The distribution and brightness of Starlink satellites is illustrated for an observer at latitude +40° at an equinox date. From left to right the Sun's distance below the western horizon is 12° (end of nautical twilight), 18° (end of astronomical twilight) and 30°.

Table C-5-β. Latitude 50°, solar declination 0.0°

```
Latitude   50.0,   solar az. 284.7,   solar el. -12.0,   solar dec.    0.0
                --------------------------- Quadrant ---------------------------
Magni-     Northwest      Southwest      Southeast      Northeast      All Quadrants
tude       30+    30-     30+    30-     30+    30-     30+    30-     30+    30-    Any
 < 5        2      3       1      2       7      3       4      2       14     10     24
5 - 6       9     23       8     14       6     11      10      2       33     50     83
6 - 7       6     25       5     38       0     12       1      8       12     83     95
 > 7        0     12       0     79       0     19       1     15        1    125    126
Sunlit     17     63      14    133      13     45      16     27       60    268    328
Shadow      0      0       0      0       0     91       0     38        0    129    129
All        17     63      14    133      13    136      16     65       60    397    457

Latitude   50.0,   solar az. 292.8,   solar el. -18.0,   solar dec.    0.0
                --------------------------- Quadrant ---------------------------
Magni-     Northwest      Southwest      Southeast      Northeast      All Quadrants
tude       30+    30-     30+    30-     30+    30-     30+    30-     30+    30-    Any
 < 5        1      2       0      1       3      0       1      0        5      3      8
5 - 6      11     16       6      4       2      0       7      0       26     20     46
6 - 7       5     25       8     35       0      0       3      1       16     61     77
 > 7        0     20       0     76       0      0       0     13        0    109    109
Sunlit     17     63      14    116       5      0      11     14       47    193    240
Shadow      0      0       0     17       8    136       5     51       13    204    217
All        17     63      14    133      13    136      16     65       60    397    457

Latitude   50.0,   solar az. 313.5,   solar el. -30.0,   solar dec.    0.0
                --------------------------- Quadrant ---------------------------
Magni-     Northwest      Southwest      Southeast      Northeast      All Quadrants
tude       30+    30-     30+    30-     30+    30-     30+    30-     30+    30-    Any
 < 5        0      0       0      0       0      0       0      0        0      0      0
5 - 6       0      0       0      0       0      0       0      0        0      0      0
6 - 7       2     12       0      1       0      0       0      0        2     13     15
 > 7        0     41       0     16       0      0       0      8        0     65     65
Sunlit      2     53       0     17       0      0       0      8        2     78     80
Shadow     15     10      14    116      13    136      16     57       58    319    377
All        17     63      14    133      13    136      16     65       60    397    457
```

Appendix C-5-γ. Latitude 50°, solar declination +23.4°

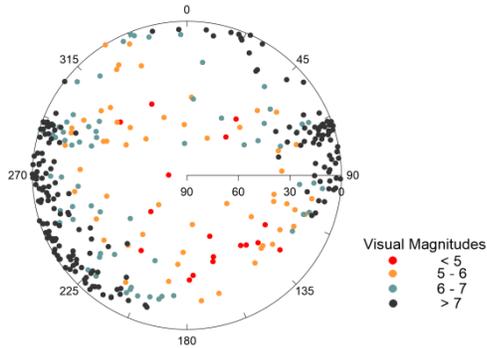

Figure C-5-γ. The distribution and brightness of Starlink satellites is illustrated for an observer at latitude +40° at the summer solstice. The Sun's distance below the northwestern horizon is 12° (end of nautical twilight). The Sun does not descend to 18° (end of astronomical twilight) or 30° below the horizon at summer solstice at this latitude.

Table C-5-γ. Latitude 50°, solar declination +23.4°

```
Latitude   50.0,  solar az. 332.2,  solar el. -12.0,  solar dec.  23.4
         --------------------------- Quadrant ---------------------------
Magni-   Northwest     Southwest     Southeast     Northeast    All Quadrants
tude     30+   30-     30+   30-     30+   30-     30+   30-    30+   30-   Any
 < 5      3     0       2     0       8     2       2     0     15     2    17
 5 - 6    9    11       7     4       5    12       6     2     27    29    56
 6 - 7    5    18       4    21       0     6       6     6     15    51    66
 > 7      0    34       1    80       0    13       2    54      3   181   184
Sunlit   17    63      14   105      13    33      16    62     60   263   323
Shadow    0     0       0    28       0   103       0     3      0   134   134
All      17    63      14   133      13   136      16    65     60   397   457
```

Appendix C-6-α. Latitude 60°, solar declination -23.4°

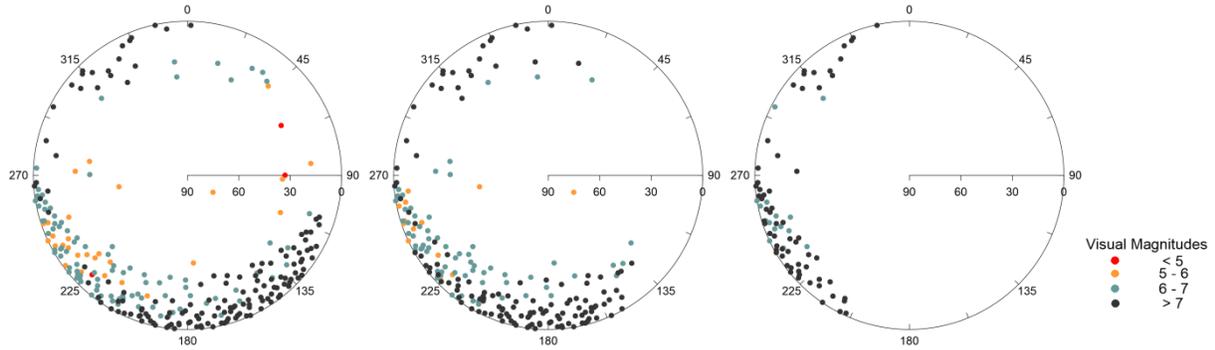

Figure C-6-α. The distribution and brightness of Starlink satellites is illustrated for an observer at latitude +60° at the winter solstice. From left to right the Sun's distance below the western horizon is 12° (end of nautical twilight), 18° (end of astronomical twilight) and 30°.

Table C-6-α. Latitude 60°, solar declination -23.4°

```
Latitude   60.0,   solar az. 243.6,   solar el. -12.0,   solar dec. -23.4
         ----------------------------- Quadrant ----------------------------
Magni-   Northwest      Southwest      Southeast      Northeast      All Quadrants
tude     30+    30-     30+    30-     30+    30-     30+    30-     30+    30-    Any
 < 5      0      0       0      1       0      0       1      1       1      2      3
 5 - 6    1      1       1     19       4      0       0      2       6     22     28
 6 - 7    2      3       0     70       0     12       0      5       2     90     92
 > 7      0     19       0     39       0     86       0      1       0    145    145
Sunlit    3     23       1    129       4     98       1      9       9    259    268
Shadow    0      0       0      0       0     27       0     16       0     43     43
All       3     23       1    129       4    125       1     25       9    302    311

Latitude   60.0,   solar az. 254.2,   solar el. -18.0,   solar dec. -23.4
         ----------------------------- Quadrant ----------------------------
Magni-   Northwest      Southwest      Southeast      Northeast      All Quadrants
tude     30+    30-     30+    30-     30+    30-     30+    30-     30+    30-    Any
 < 5      0      0       0      0       0      0       0      0       0      0      0
 5 - 6    0      0       1     11       1      0       0      0       2     11     13
 6 - 7    3      2       0     53       1      6       0      1       4     62     66
 > 7      0     21       0     65       0     42       0      2       0    130    130
Sunlit    3     23       1    129       2     48       0      3       6    203    209
Shadow    0      0       0      0       2     77       1     22       3     99    102
All       3     23       1    129       4    125       1     25       9    302    311

Latitude   60.0,   solar az. 274.8,   solar el. -30.0,   solar dec. -23.4
         ----------------------------- Quadrant ----------------------------
Magni-   Northwest      Southwest      Southeast      Northeast      All Quadrants
tude     30+    30-     30+    30-     30+    30-     30+    30-     30+    30-    Any
 < 5      0      0       0      0       0      0       0      0       0      0      0
 5 - 6    0      0       0      0       0      0       0      0       0      0      0
 6 - 7    0      2       0     13       0      0       0      0       0     15     15
 > 7      0     17       0     45       0      0       0      0       0     62     62
Sunlit    0     19       0     58       0      0       0      0       0     77     77
Shadow    3      4       1     71       4    125       1     25       9    225    234
All       3     23       1    129       4    125       1     25       9    302    311
```

Appendix C-6-β. Latitude 60°, solar declination 0.0°

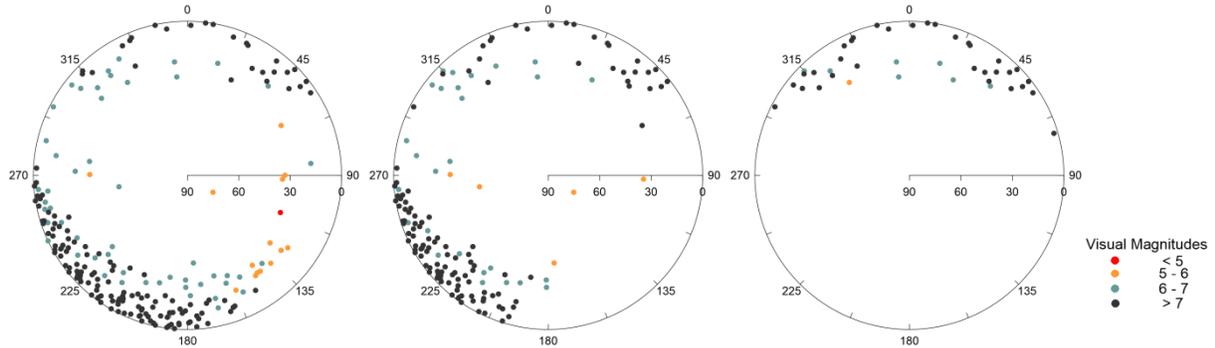

Figure C-6-β. The distribution and brightness of Starlink satellites is illustrated for an observer at latitude +40° at an equinox date. From left to right the Sun's distance below the western horizon is 12° (end of nautical twilight), 18° (end of astronomical twilight) and 30°.

Table C-6-β. Latitude 60°, solar declination 0.0°

```
Latitude   60.0,   solar az. 291.6,   solar el. -12.0,   solar dec.   0.0
                    ---------------------------- Quadrant ----------------------------
Magni-     Northwest      Southwest      Southeast      Northeast      All Quadrants
tude       30+    30-     30+    30-    30+    30-     30+    30-     30+    30-    Any
 < 5        0      0       0      0      1      0       0      0       1      0      1
 5 - 6      1      0       0      0      2      9       1      1       4     10     14
 6 - 7      2     13       1     23      1     10       0      3       4     49     53
 > 7        0     10       0    106      0     15       0     18       0    149    149
Sunlit      3     23       1    129      4     34       1     22       9    208    217
Shadow      0      0       0      0      0     91       0      3       0     94     94
All         3     23       1    129      4    125       1     25       9    302    311

Latitude   60.0,   solar az. 304.2,   solar el. -18.0,   solar dec.   0.0
                    ---------------------------- Quadrant ----------------------------
Magni-     Northwest      Southwest      Southeast      Northeast      All Quadrants
tude       30+    30-     30+    30-    30+    30-     30+    30-     30+    30-    Any
 < 5        0      0       0      0      0      0       0      0       0      0      0
 5 - 6      1      0       1      0      3      0       0      0       5      0      5
 6 - 7      2     12       0     12      0      0       0      1       2     25     27
 > 7        0     11       0     86      0      0       0     19       0    116    116
Sunlit      3     23       1     98      3      0       0     20       7    141    148
Shadow      0      0       0     31      1    125       1      5       2    161    163
All         3     23       1    129      4    125       1     25       9    302    311

Latitude   60.0,   solar az.   0.0,   solar el. -30.0,   solar dec.   0.0
                    ---------------------------- Quadrant ----------------------------
Magni-     Northwest      Southwest      Southeast      Northeast      All Quadrants
tude       30+    30-     30+    30-    30+    30-     30+    30-     30+    30-    Any
 < 5        0      0       0      0      0      0       0      0       0      0      0
 5 - 6      0      1       0      0      0      0       0      0       0      1      1
 6 - 7      1      3       0      0      0      0       0      3       1      6      7
 > 7        0     14       0      0      0      0       0     18       0     32     32
Sunlit      1     18       0      0      0      0       0     21       1     39     40
Shadow      2      5       1    129      4    125       1      4       8    263    271
All         3     23       1    129      4    125       1     25       9    302    311
```

Appendix C-6-γ. Latitude 60°, solar declination +23.4°

The Sun does not descend to 12° (end of nautical twilight) below the horizon at summer solstice at this latitude.